\documentclass[amsmath,amssymb,nofootinbib,notitlepage,superscriptaddress,twocolumn]{revtex4-1}
\pdfoutput=1

\usepackage{aas_macros,esint,graphicx}
\usepackage[bottom]{footmisc}
\usepackage[colorlinks=true]{hyperref}
\let\originalleft\left
\let\originalright\right
\renewcommand{\left}{\mathopen{}\mathclose\bgroup\originalleft}
\renewcommand{\right}{\aftergroup\egroup\originalright}

\newcommand{\ab}[1]{\left|#1\right|}
\newcommand{\br}[1]{\left[#1\right]}
\newcommand{\cu}[1]{\left\{#1\right\}}
\newcommand{\pa}[1]{\left(#1\right)}

\newcommand{\ed}{\mathop{}\!\mathrm{d}}
\newcommand{\pd}{\mathop{}\!\partial}

\DeclareMathOperator\am{am}
\DeclareMathOperator\cn{cn}
\DeclareMathOperator\dn{dn}
\DeclareMathOperator\sn{sn}
\DeclareMathOperator\tn{sc}
\DeclareMathOperator\sign{sign}

\begin{document}

\title{Null Geodesics of the Kerr Exterior}

\author{Samuel E. Gralla}
\email{sgralla@email.arizona.edu}
\affiliation{Department of Physics, University of Arizona, Tucson, Arizona 85721, USA}
\author{Alexandru Lupsasca}
\email{lupsasca@fas.harvard.edu}
\affiliation{Center for the Fundamental Laws of Nature, Harvard University, Cambridge, Massachusetts 02138, USA}
\affiliation{Society of Fellows, Harvard University, Cambridge, Massachusetts 02138, USA}

\begin{abstract}
The null geodesic equation in the Kerr spacetime can be expressed as a set of integral equations involving certain potentials.  We classify the roots of these potentials and express the integrals in manifestly real Legendre elliptic form.  We then solve the equations using Jacobi elliptic functions, providing the complete set of null geodesics of the Kerr exterior as explicit parameterized curves.
\end{abstract}

\maketitle

\section{Introduction}

Null geodesics form perhaps the most important structure possessed by a Lorentzian spacetime.  The study of the null geodesic equation in the Kerr geometry began in 1968 with the seminal work of Carter \cite{Carter1968}, who used the separability of its Hamilton-Jacobi formulation to reduce it to quadratures.  Bardeen \cite{Bardeen1973} initiated the detailed study of its solution space, which has now been mapped out in impressive detail \cite{ONeill1995,Chandrasekhar1983,Hackmann2010}.  Many of the relevant integrals have previously been expressed in elliptic form (notably in Refs.~\cite{Rauch1994,Dexter2009,Kapec2019}), and parameterized solutions using the Weierstrass elliptic function were given in Ref.~\cite{Hackmann2010}.

In this paper, we revisit the problem of the classification and solution of Kerr null geodesics with the goals of \textit{completeness} and \textit{convenience}.  Our results are complete in that every finite-measure case is considered, and we give both  ``integral solutions'' (analytic expressions for the fundamental integrals) as well as explicit parameterized trajectories.\footnote{We do limit our discussion to the Kerr exterior, however.}  Moreover, our results are convenient in that: (1) all expressions are manifestly real, with no ``canceling'' internal imaginary parts; (2) all trajectories are fully explicit, with no need to solve auxiliary equations or glue together different solutions at turning points; (3) the parameterized solutions display the initial conditions explicitly; and (4) the use of special functions is limited to the elliptic integrals and Jacobi elliptic functions first defined two centuries ago.  While previous approaches achieve one or more of these goals, we are unaware of any previous work that simultaneously attains them all.

Our own interest in this problem was kindled by the need to understand astronomical observations \cite{EHT2019a}, but we hope that the results presented herein will find a wider range of applications.  Ideally, the Kerr afficionado will learn something about the general structure of the null geodesics, while the busy physicist or astronomer can obtain quick answers to definite questions about the propagation of light around a rotating black hole.

This paper is organized as follows.  In Sec.~\ref{sec:Approach}, we introduce the basic formulas and explain our general approach.  In Sec.~\ref{sec:AngularPotential}, we classify and solve for the different types of polar motion, before doing the same for the radial motion in Sec.~\ref{sec:RadialPotential} and App.~\ref{app:RadialIntegrals}.  We then compare to previous work in Sec.~\ref{sec:Comparison}.  Finally, in Sec.~\ref{sec:Summary}, we summarize our results and explain how to use them in practice.

\section{General approach}
\label{sec:Approach}

We work with Boyer-Lindquist coordinates $(t,r,\theta,\phi)$ on the spacetime of a Kerr black hole with mass $M$ and angular momentum $J=Ma$, and define
\begin{align}
	\Sigma(r,\theta)=r^2+a^2\cos^2{\theta},\quad
	\Delta(r)=r^2-2Mr+a^2.
\end{align}
The roots of $\Delta(r)$ correspond to the outer/inner horizons
\begin{align}
	r_\pm=M\pm\sqrt{M^2-a^2}.
\end{align}
We assume $0<a<M$ throughout the paper.  Taking the nonrotating limit $a\to0$ is generally straightforward, whereas the extremal limit $a\to M$ presents some subtleties that we defer to future work.

Let $p^\mu$ denote the four-momentum of a photon, with $p^t>0$ providing the time orientation.  The trajectory possesses three conserved quantities,
\begin{align}
	\label{eq:ConservedQuantities}
	E&=-p_t,\quad
	L=p_\phi,\\
	Q&=p_\theta^2-\cos^2{\theta}\pa{a^2p_t^2-p_\phi^2\csc^2{\theta}},
\end{align}
corresponding to the energy at infinity,\footnote{We exclude the measure-zero set of geodesics with $E=0$ exactly.} angular momentum about the spin axis, and Carter integral, respectively.  Only the sign of the energy has physical meaning, so it is convenient to work with energy-rescaled quantities
\begin{align}
	\label{eq:RescaledConservedQuantities}
	\lambda=\frac{L}{E},\quad
	\eta=\frac{Q}{E^2}.
\end{align}
The four-momentum $p^\mu$ can then be reconstructed as
\begin{subequations}
\label{eq:GeodesicEquation}
\begin{align}
	\frac{\Sigma}{E}p^r&=\pm_r\sqrt{\mathcal{R}(r)},\\
	\frac{\Sigma}{E}p^\theta&=\pm_\theta\sqrt{\Theta(\theta)},\\
	\frac{\Sigma}{E}p^\phi&=\frac{a}{\Delta}\pa{r^2+a^2-a\lambda}+\frac{\lambda}{\sin^2{\theta}}-a,\\
	\frac{\Sigma}{E}p^t&=\frac{r^2+a^2}{\Delta}\pa{r^2+a^2-a\lambda}+a\pa{\lambda-a\sin^2{\theta}},
\end{align}
\end{subequations}
where we introduced ``potentials''
\begin{align}
	\label{eq:RadialPotential}
	\mathcal{R}(r)&=\pa{r^2+a^2-a\lambda}^2-\Delta(r)\br{\eta+\pa{\lambda-a}^2},\\
	\label{eq:AngularPotential}
	\Theta(\theta)&=\eta+a^2\cos^2{\theta}-\lambda^2\cot^2{\theta}.
\end{align}
The symbols $\pm_r$ and $\pm_\theta$ indicate the sign of $p^r$ and $p^\theta$, respectively.  Turning points in the $r$ and $\theta$ motions occur at zeros of the radial and angular potentials $\mathcal{R}(r)$ and $\Theta(\theta)$, respectively.

There are two closely related ways to proceed with the solution of these equations.  The first is to introduce a new parameter, the ``Mino time'' $\tau$ \cite{Mino2003}, defined by\footnote{The geodesic $x^\mu(\tau)$ is future/past-directed according to whether $E$ is positive/negative.  Sending $\tau\to-\tau$ reverses the future/past direction of the parameterized curve $x^\mu(\tau)$.}
\begin{align}
	\frac{dx^\mu}{d\tau}=\frac{\Sigma}{E}p^\mu.
\end{align}
This method converts Eqs.~\eqref{eq:GeodesicEquation} into four decoupled ordinary differential equations for $x^\mu(\tau)$.  Alternatively, one may also convert the equations into integral form,
\begin{align}
	I_r&=G_\theta,\\
	\label{eq:Deltaphi}
	\phi_o-\phi_s&=I_\phi+\lambda G_\phi,\\
	\label{eq:Deltat}
	t_o-t_s&=I_t+a^2G_t,
\end{align}
where $x^\mu_s$ and $x^\mu_o$ are ``source'' and ``observer'' points, $\phi$ can take any real value (with $\lfloor(\phi_o-\phi_s)/(2\pi)\rfloor$ the number of azimuthal windings of the trajectory), and we define
\begin{subequations}
\label{eq:PathIntegrals}
\begin{align}
	\label{eq:Ir}
	I_r&=\fint_{r_s}^{r_o}\frac{\ed r}{\pm_r\sqrt{\mathcal{R}(r)}},\\
	I_\phi&=\fint_{r_s}^{r_o}\frac{a\pa{2Mr-a\lambda}}{\pm_r\Delta(r)\sqrt{\mathcal{R}(r)}}\ed r,\\
	I_t&=\fint_{r_s}^{r_o}\frac{r^2\Delta(r)+2Mr\pa{r^2+a^2-a\lambda}}{\pm_r\Delta(r)\sqrt{\mathcal{R}(r)}}\ed r,\\
	G_\theta&=\fint_{\theta_s}^{\theta_o}\frac{\ed\theta}{\pm_\theta\sqrt{\Theta(\theta)}},\\
	\label{eq:Gphi}
	G_\phi&=\fint_{\theta_s}^{\theta_o}\frac{\csc^2{\theta}}{\pm_\theta\sqrt{\Theta(\theta)}}\ed\theta,\\
	\label{eq:Gt}
	G_t&=\fint_{\theta_s}^{\theta_o}\frac{\cos^2{\theta}}{\pm_\theta\sqrt{\Theta(\theta)}}\ed\theta.
\end{align}
\end{subequations}
Here, the slash notation $\fint$ indicates that these integrals are to be understood as path integrals along the trajectory connecting $x^\mu_s$ and $x^\mu_o$, such that the signs $\pm_r$ and $\pm_\theta$ switch at radial and angular turning points, respectively.  In particular, all the integrals $I_i$ and $G_i$ are monotonically increasing along the trajectory.

These two approaches are related by the fact that $I_r$ and $G_\theta$ are both equal to the elapsed Mino time,\footnote{$\tau$ is also related to the fractional number of orbits executed \cite{KerrLensing}.}
\begin{align}
	\tau=I_r=G_\theta,
\end{align}
where we set $\tau=0$ at the source point $x^\mu_s$.  The Mino time approach is more convenient for analyzing individual trajectories, while the integral approach is more useful for determining general properties.

The elapsed affine time (satisfying $dx^\mu/d\sigma=p^\mu$) is
\begin{align}
	\sigma_o-\sigma_s=I_\sigma+a^2G_t,
\end{align}
where
\begin{align}
	\label{eq:Isigma}
	I_\sigma=\fint_{r_s}^{r_o}\frac{r^2}{\pm_r\sqrt{\mathcal{R}(r)}}\ed r.
\end{align}

Our main results are as follows.  First, we systematically classify the roots of the radial and angular potentials, and thereby determine the allowed ranges of the $r$ and $\theta$ motion as a function of the conserved quantities $(\lambda,\eta)$.  Then, for each of the cases that may arise, and for each integral $I_i$ or $G_i$, we find an antiderivative that is real and smooth over the relevant range of $r$ or $\theta$.  All of our antiderivatives are reduced to manifestly real Legendre elliptic form.  That is, they are expressed in terms of the (incomplete) elliptic integrals $F(\varphi|k)$, $E(\varphi|k)$, and $\Pi(n;\varphi|k)$ of the first, second, and third kind, respectively, which are real and smooth provided $\max(k,n)<1$.\footnote{Our conventions are listed in App.~A of Ref.~\cite{Kapec2019}; these also match the built-in implementation in \textit{Mathematica 12}.}  When $\varphi=\pi/2$, the integrals become ``complete'' and are denoted by $K(k)=F(\pi/2|k)$, or else by $\Pi(n;k)$ and $E(k)$ with the first argument $\varphi$ omitted.

Our notation for antiderivatives will be a calligraphic version of the original symbol, and we will choose the plus sign in the integrand.  For example, the antiderivative $\mathcal{I}_r$ associated with $I_r$ in Eq.~\eqref{eq:Ir} will satisfy
\begin{align}
	\frac{d\mathcal{I}_r}{dr}=\frac{1}{\sqrt{\mathcal{R}(r)}}.
\end{align}
These antiderivatives are useful for both the Mino-time approach and the integral approach.  For the Mino-time approach, we invert the integrals to provide full parameterized trajectories $x^\mu(\tau)$ in terms of the initial data (initial position $x^\mu_s$ as well as the initial signs $\pm_r$ and $\pm_\theta$).\footnote{The full initial derivative can then be reconstructed from Eqs.~\eqref{eq:GeodesicEquation}, showing how this initial value problem is equivalent to the original second-order initial value problem for the geodesic equation.}  For the integral approach, we provide formulas that give each of the path integrals \eqref{eq:PathIntegrals} as a function of the initial position, initial sign of momentum, final position, and  number of turning points.

\section{Angular potential and integrals}
\label{sec:AngularPotential}

We assume that $0<\theta<\pi$ to avoid the singularities of the spherical coordinate system.  In terms of $u=\cos^2\theta$, the angular potential \eqref{eq:AngularPotential} is given by
\begin{align}
	\pa{1-u}\Theta(u)&=\eta+\pa{a^2-\eta-\lambda^2}u-a^2u^2.
\end{align}
The right-hand side is a quadratic polynomial, whose roots $u_\pm$ are given by
\begin{align}
	u_\pm=\triangle_\theta\pm\sqrt{\triangle_\theta^2+\frac{\eta}{a^2}},\quad
	\triangle_\theta=\frac{1}{2}\pa{1-\frac{\eta+\lambda^2}{a^2}}.
\end{align}
The four roots of $\Theta(\theta)$ are thus $\arccos\pa{\pm\sqrt{u_\pm}}$, or
\begin{align}
	\theta_1&=\arccos\pa{\sqrt{u_+}},\\
	\theta_2&=\arccos\pa{\sqrt{u_-}},\\
	\theta_3&=\arccos\pa{-\sqrt{u_-}},\\
	\theta_4&=\arccos\pa{-\sqrt{u_+}}.
\end{align}
Roots coincide when (and only when) $u_+=0$, $u_-=0$, or $u_+=u_-$.  These conditions define curves through the $(\lambda,\eta)$-plane that divide it up into several regions.  In each such region, the ``character'' of the potential---that is, the number of real roots and the sign of the potential on either side of them---cannot change.  As such, we may determine the character by evaluating a single point in each region.  

\begin{figure}
	\centering
	\includegraphics[width=\columnwidth]{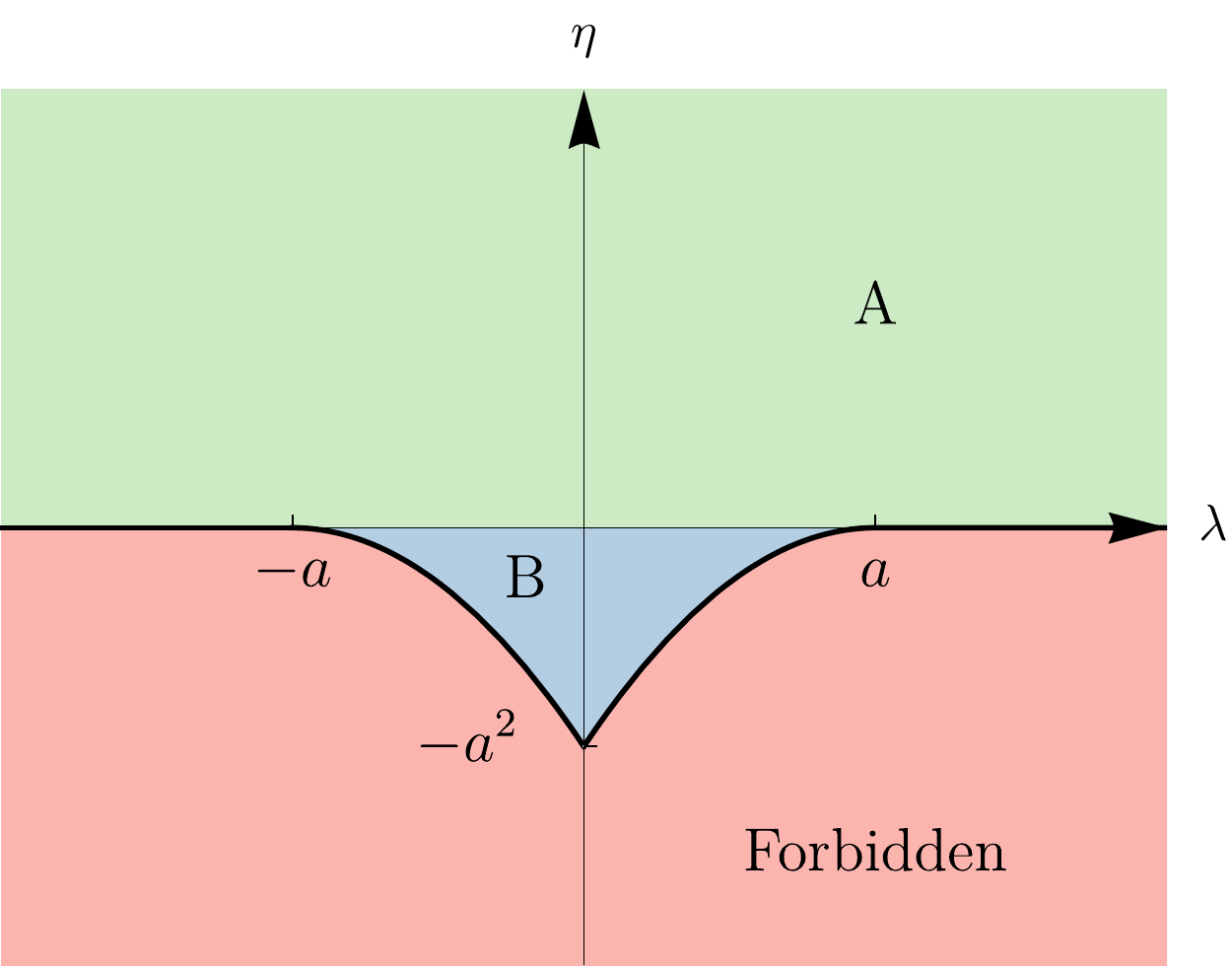}
	\caption{Regions corresponding to the two types A and B of allowed polar motion.  Vortical (type B) geodesics only exist around spinning black holes ($a>0$).}
	\label{fig:AngularRegions}
\end{figure}

Performing this exercise reveals the following structure (Fig.~\ref{fig:AngularRegions}).  Motion is allowed in the region where $0<u_+$ and $u_-<1$, and these conditions imply a lower bound
\begin{align}
	\label{eq:Protrusion}
	\eta\geq
	\begin{cases}
		0&\quad\ab{\lambda}\ge a,\\
		-\pa{\ab{\lambda}-a}^2&\quad\ab{\lambda}\le a.
	\end{cases}
\end{align}
Within this allowed region, the line $\eta=0$ of double roots delineates regions with two characters of null geodesic:
\begin{itemize}
\item[A.]
Ordinary geodesics ($\eta>0$). These admit two real roots $\theta_1<\pi/2<\theta_4$, with the potential positive in between them.  The photon librates between $\theta_1$ and $\theta_4$, crossing the equatorial plane each time.
\item[B.]
Vortical geodesics ($\eta<0$).  These admit four real roots $\theta_1<\theta_2<\pi/2<\theta_3<\theta_4$, with the potential positive in $(\theta_1,\theta_2)$ and $(\theta_3,\theta_4$).  There are two distinct motions: one that librates between turning points $(\theta_1,\theta_2)$ in the northern hemisphere, and one that librates between turning points $(\theta_3,\theta_4)$ in the southern hemisphere.
\end{itemize}
The measure-zero case $\eta=0$ contains equatorial orbits with no turning points (a limit of type A motion), as well as orbits with at most one nonequatorial turning point $\theta_{1,4}$ (a limit of type B motion, in which $\theta_{2,3}\to\pi/2$, where the angular potential develops a double root).

For the analysis below, it will be helpful to have noted that the following differential equations are satisfied:
\begin{align}
	\label{eq:AngularODE}
	\pa{\frac{d\theta_o}{d\tau}}^2&=\Theta(\theta_o(\tau)),\\
	\label{eq:GphiODE}
	\frac{dG_\phi}{d\tau}&=\csc^2\br{\theta_o(\tau)},\\
	\label{eq:GtODE}
	\frac{dG_t}{d\tau}&=\cos^2\br{\theta_o(\tau)}.
\end{align}

\subsection{Ordinary motion}

We begin with ordinary geodesics (type A with $\eta>0$), which oscillate between turning points $\theta_-<\theta_+$ given by
\begin{align}
	\theta_\pm=\arccos\pa{\mp\sqrt{u_+}},
\end{align}
so that $\theta_-=\theta_1\in\pa{0,\pi/2}$ and $\theta_+=\theta_4\in\pa{\pi/2,\pi}$.  This motion is symmetric about the equator and $\theta_+=\pi-\theta_-$.

The angular integrals $G_\theta$, $G_\phi$, and $G_t$ were reduced to manifestly real elliptic form in Ref.~\cite{Kapec2019}.  Since $u_+/u_-<0$, the antiderivatives
\begin{align}
	\label{eq:GthetaAntiderivativeA}
	\mathcal{G}_\theta&=-\frac{1}{\sqrt{-u_-a^2}}F\pa{\arcsin\pa{\frac{\cos{\theta}}{\sqrt{u_+}}}\left|\frac{u_+}{u_-}\right.},\\
	\label{eq:GphiAntiderivativeA}
	\mathcal{G}_\phi&=-\frac{1}{\sqrt{-u_-a^2}}\Pi\pa{u_+;\arcsin\pa{\frac{\cos{\theta}}{\sqrt{u_+}}}\left|\frac{u_+}{u_-}\right.},\\
	\label{eq:GtAntiderivativeA}
	\mathcal{G}_t&=\frac{2u_+}{\sqrt{-u_-a^2}}E'\pa{\arcsin\pa{\frac{\cos{\theta}}{\sqrt{u_+}}}\left|\frac{u_+}{u_-}\right.},
\end{align}
are real and smooth.  Here, we defined
\begin{align}
	E'(\varphi|k):=\pd_kE(\varphi|k)
	=\frac{E(\varphi|k)-F(\varphi|k)}{2k}.
\end{align}
We denote the values over one half-libration with a hat,
\begin{align}
	\hat{G}_\theta&=\int_{\theta_-}^{\theta_+}\frac{\ed\theta}{\sqrt{\Theta(\theta)}}
	=\frac{2}{\sqrt{-u_-a^2}}K\pa{\frac{u_+}{u_-}},\\
	\hat{G}_\phi&=\int_{\theta_-}^{\theta_+}\frac{\csc^2{\theta}}{\sqrt{\Theta(\theta)}}\ed\theta
	=\frac{2}{\sqrt{-u_-a^2}}\Pi\pa{u_+\left|\frac{u_+}{u_-} \right.},\\
	\hat{G}_t&=\int_{\theta_-}^{\theta_+}\frac{\cos^2{\theta}}{\sqrt{\Theta(\theta)}}\ed\theta
	=-\frac{4u_+}{\sqrt{-u_-a^2}}E'\pa{\frac{u_+}{u_-}}.
\end{align}
Note that since $G_\theta=\tau$ is the Mino time, the polar motion $\theta_o(\tau)$ has a Mino-time period of $2\hat{G}_\theta$.

\subsubsection{Inversion for \texorpdfstring{$\theta_o(\tau)$}{polar motion}}

Now consider the path integral $G_\theta=\tau$ beginning from $\theta=\theta_s$.  Before the first turning point is reached, we have 
\begin{align}
	\label{eq:AngularMinoTime}
	\tau=G_\theta
	=\nu_\theta\pa{\mathcal{G}_\theta^o-\mathcal{G}_\theta^s},\quad
	\nu_\theta=\sign\pa{p_s^\theta},
\end{align}
where $\mathcal{G}_\theta^i$ indicates the antiderivative $\mathcal{G}_\theta$ evaluated at the source ($i=s$) or observer ($i=o$).  This equation can be inverted for $\theta_o$ using the Jacobi elliptic sine function, 
\begin{align}
	\label{eq:JacobiSN}
	\sn\pa{F(\arcsin{\varphi}|k)}=\varphi,
\end{align}
which is odd in its first argument, $\sn(-\varphi|k)=-\sn(\varphi|k)$.  Combining Eqs.~\eqref{eq:GthetaAntiderivativeA}, \eqref{eq:AngularMinoTime}, and \eqref{eq:JacobiSN} therefore gives
\begin{align}
	\label{eq:InversionA}
	\frac{\cos{\theta_o}}{\sqrt{u_+}}=-\nu_\theta\sn\pa{\sqrt{-u_-a^2}\pa{\tau+\nu_\theta\mathcal{G}_\theta^s}\left|\frac{u_+}{u_-}\right.}.
\end{align}
Although Eq.~\eqref{eq:InversionA} was derived under the assumption that a turning point has not yet been reached, it in fact continues properly through turning points to provide the full parameterized trajectory $\theta_o(\tau)$, as follows.  Noting that $\sn(\varphi|k)$ oscillates smoothly between $-1$ and $+1$ with half-period $2K(k)$, we see that Eq.~\eqref{eq:InversionA} defines a trajectory $\theta_o(\tau)$ that oscillates between $\theta_-$ and $\theta_+$ with half-period $\hat{G}_\theta$.  Thus, it has the correct quantitative behavior at turning points, and we need only check that it satisfies the squared differential equation \eqref{eq:AngularODE}, which is easily verified using the elliptic identities \cite{Abramowitz1972}
\begin{align}
	\label{eq:JacobiSNCN}
	\cn^2(\varphi|k)+\sn^2(\varphi|k)&=1,\\
	\label{eq:JacobiSNDN}
	\dn^2(\varphi|k)+k\sn^2(\varphi|k)&=1.
\end{align}
This completes the proof that Eq.~\eqref{eq:InversionA} is the unique solution for $\theta_o(\tau)$ with initial conditions $\theta_o(0)=\theta_s$ and $\sign\br{\theta_o'(0)}=\nu_\theta$.

\subsubsection{Path integrals as functions of Mino time}

We have $G_\theta=\tau$ by definition, and the other path integrals may be expressed in terms of Mino time $\tau$ as follows.  Before a turning point is reached, we have
\begin{align}
	\label{eq:DefiniteGphi}
	G_\phi&=\nu_\theta\pa{\mathcal{G}_\phi^o-\mathcal{G}_\phi^s},\\
	\label{eq:DefiniteGt}
	G_t&=\nu_\theta\pa{\mathcal{G}_t^o-\mathcal{G}_t^s}.
\end{align}
To manipulate these equations, we will invoke a second inversion formula,
\begin{align}
	\label{eq:JacobiAmplitude}
	\am(\varphi|k)=\arcsin\pa{\sn(\varphi|k)},\quad
	\ab{\varphi}\le K(k).
\end{align}
where the Jacobi amplitude $\am(\varphi|k)$ is defined as the inverse of the elliptic integral of the first kind $F(\varphi|k)$,
\begin{align}
	F(\am(\varphi|k)|k)=\varphi.
\end{align}
Applying the formula \eqref{eq:JacobiAmplitude} to Eq.~\eqref{eq:InversionA} yields
\begin{align}
	\label{eq:UnfoldingA}
	\arcsin\pa{\frac{\cos{\theta_o}}{\sqrt{u_+}}}=-\nu_\theta\Psi_\tau,
\end{align}
where the (monotonically increasing in $\tau$) amplitude is
\begin{align}
	\Psi_\tau=\am\pa{\sqrt{-u_-a^2}\pa{\tau+\nu_\theta\mathcal{G}_\theta^s}\left|\frac{u_+}{u_-}\right.},
\end{align}
and the restriction $|\varphi|\le K(k)$ is satisfied on account of our assumption that a turning point has not yet been reached.  Plugging Eq.~\eqref{eq:UnfoldingA} into Eqs.~\eqref{eq:DefiniteGphi} and \eqref{eq:DefiniteGt} as needed, and noting that both $\Pi(n;-\varphi|k)=-\Pi(n;\varphi|k)$ and $E'(-\varphi|k)=-E'(\varphi|k)$ are odd in $\varphi$, we then find
\begin{align}
	\label{eq:GphiIntegralA}
	G_\phi&=\frac{1}{\sqrt{-u_-a^2}}\Pi\pa{u_+;\Psi_\tau\left|\frac{u_+}{u_-}\right.}-\nu_\theta\mathcal{G}_\phi^s,\\
	\label{eq:GtIntegralA}
	G_t&=-\frac{2u_+}{\sqrt{-u_-a^2}}E'\pa{\Psi_\tau\left|\frac{u_+}{u_-}\right.}-\nu_\theta\mathcal{G}_t^s.
\end{align}
Although Eqs.~\eqref{eq:GphiIntegralA} and \eqref{eq:GtIntegralA} were derived under the assumption that a turning point does not occur, they in fact extend properly through turning points to give the complete path integrals $G_\phi$ and $G_t$, as follows.  Since $\am(\varphi|k)$, $\Pi(n;\varphi|k)$ and $E'(\varphi|k)$ are real and smooth functions of $\varphi$ provided $\max(k,n)<1$ (satisfied here for $k=u_+/u_-$ and $n=u_+$), it follows that the candidate formulas for $G_\phi$ and $G_t$ are real and smooth.  Thus, we need only check the differential equations \eqref{eq:GphiODE} and \eqref{eq:GtODE}, which is straightforward using the identity \eqref{eq:JacobiSNDN}.  This completes the proof that Eqs.~\eqref{eq:GphiIntegralA} and \eqref{eq:GtIntegralA} give the full path integrals \eqref{eq:Gphi} and \eqref{eq:Gt}.

Notice that Eq.~\eqref{eq:InversionA} may also be put in a similar form using $\sn(\varphi|k)=\sin\pa{\am(\varphi|k)}$,
\begin{align}
	\label{eq:InversionAmplitudeA}
	\cos{\theta_o(\tau)}=-\nu_\theta\sqrt{u_+}\sin{\Psi_\tau}.
\end{align}

\subsubsection{Path integrals in terms of turning points}

Finally, it is useful for some purposes to express the path integrals $G_i$ as functions of $\theta_s$, $\theta_o$, $\nu_\theta$, and the number of turning points $m$ encountered along the trajectory.  A general treatment is given in App.~\ref{app:UnpackingPathIntegrals} below.  For type A motion, the antiderivatives $\mathcal{G}_i$ are odd under interchange of $+$ and $-$, 
\begin{align}
	\mathcal{G}_i^+=-\mathcal{G}_i^-,
\end{align}
where the $\pm$ denotes evaluation of $\mathcal{G}_i$ at $\theta=\theta_\pm$.  This property originates from the equatorial reflection symmetry $\Theta(\theta)=\Theta(\pi-\theta)$ of the angular potential.  The general result \eqref{eq:CombinedH} therefore reduces to
\begin{align}
	\label{eq:OrdinaryPathIntegrals}
	G_i=m\hat{G}_i+\nu_\theta\br{(-1)^m\mathcal{G}_i^o-\mathcal{G}_i^s},
\end{align}
for $i\in\cu{\theta,\phi,t}$, in agreement with Eqs.~(80) in Ref.~\cite{Kapec2019}. 

It is instructive to examine the relationship between this formula and the above expressions parameterized by Mino time.  When $i=\theta$, Eq.~\eqref{eq:OrdinaryPathIntegrals} is the Mino time itself,
\begin{align}
	\label{eq:UnpackedTau}
	\tau=m\hat{G}_\theta+\nu_\theta\br{(-1)^m\mathcal{G}_\theta^o-\mathcal{G}_\theta^s}. 
\end{align}
By using the quasiperiodicity properties
\begin{subequations}
\label{eq:QuasiPeriodicity}
\begin{align}
	\am(\varphi+2K(k)|k)&=\am(\varphi|k)+\pi,\quad
	k<1,\\
	F(\varphi+\pi|k)&=F(\varphi|k)+2K(k),\\
	\Pi(n;\varphi+\pi|k)&=\Pi(n;\varphi|k)+2\Pi(n;k),\\
	E(\varphi+\pi|k)&=E(\varphi|k)+2E(k),
\end{align}
\end{subequations}
one can plug Eq.~\eqref{eq:UnpackedTau} into Eqs.~\eqref{eq:GphiIntegralA}, \eqref{eq:GtIntegralA}, and \eqref{eq:InversionAmplitudeA} to recover Eq.~\eqref{eq:OrdinaryPathIntegrals} for $i\in\cu{\phi,t}$, and verify that $\theta_o(\tau)=\theta_o$, as required for consistency.

\subsection{Vortical motion}

We next turn to vortical geodesics (type B with $\eta<0$), which oscillate within a single hemisphere determined by
\begin{align}
	\label{eq:h}
	h=\sign\pa{\cos{\theta}}.
\end{align}
The motion lies within a cone $\theta_-<\theta_+$ in the northern hemisphere ($h=+1$), or $\theta_+<\theta_-$ in the southern hemisphere ($h=-1$), with the turning points given by
\begin{align}
	\theta_\pm=\arccos\pa{h\sqrt{u_\mp}},
\end{align}
so that $\theta_{-,+}=\theta_{1,2}$ for $h=1$ and $\theta_{+,-}=\theta_{3,4}$ for $h=-1$.

The angular integrals $G_\theta$, $G_\phi$, and $G_t$ were reduced to manifestly real elliptic form in Ref.~\cite{Kapec2019}.  Since $u_+/u_->0$, the antiderivatives
\begin{align}
	\label{eq:GthetaAntiderivativeB}
	\mathcal{G}_\theta&=-\frac{h}{\sqrt{u_-a^2}}F\pa{\Upsilon\left|1-\frac{u_+}{u_-}\right.},\\
	\label{eq:GphiAntiderivativeB}
	\mathcal{G}_\phi&=-\frac{h}{\pa{1-u_-}\sqrt{u_-a^2}}\Pi\pa{\frac{u_+-u_-}{1-u_-};\Upsilon\left|1-\frac{u_+}{u_-}\right.},\\
	\label{eq:GtAntiderivativeB}
	\mathcal{G}_t&=-h\sqrt{\frac{u_-}{a^2}}E\pa{\Upsilon\left|1-\frac{u_+}{u_-}\right.},
\end{align}
are real and smooth, with
\begin{align}
	\Upsilon=\arcsin\sqrt{\frac{\cos^2{\theta}-u_-}{u_+-u_-}}.
\end{align}
Their values over one half-libration are
\begin{align}
	\hat{G}_\theta&=h\int_{\theta_-}^{\theta_+}\frac{\ed\theta}{\sqrt{\Theta(\theta)}}
	=\frac{1}{\sqrt{u_-a^2}}K\pa{1-\frac{u_+}{u_-}},\\
	\hat{G}_\phi&=h\int_{\theta_-}^{\theta_+}\frac{\csc^2{\theta}}{\sqrt{\Theta(\theta)}}\ed \theta\nonumber\\
	&=\frac{1}{\pa{1-u_-}\sqrt{u_-a^2}}\Pi\pa{\frac{u_+-u_-}{1-u_-};1-\frac{u_+}{u_-}},\\
	\hat{G}_t&=h\int_{\theta_-}^{\theta_+}\frac{\cos^2{\theta}}{\sqrt{\Theta(\theta)}}\ed\theta
	=\sqrt{\frac{u_-}{a^2}}E\pa{1-\frac{u_+}{u_-}},
\end{align}
and $\hat{G}_\theta$ once again denotes the Mino-time half-period of the polar motion $\theta_o(\tau)$.

\subsubsection{Inversion for \texorpdfstring{$\theta_o(\tau)$}{polar motion}}

Now consider the path integral $G_\theta=\tau$ beginning from $\theta=\theta_s$.  Before the first turning point is reached, we once again have Eq.~\eqref{eq:AngularMinoTime}, which can yet again be inverted using Eq.~\eqref{eq:JacobiSN} to obtain
\begin{align}
	\label{eq:AlmostInversionB}
	\sqrt{\frac{\cos^2{\theta_o}-u_-}{u_+-u_-}}&=-h\nu_\theta\sn\pa{\sqrt{u_-a^2}\pa{\tau+\nu_\theta\mathcal{G}_\theta^s}\left|1-\frac{u_+}{u_-}\right.}.
\end{align}
Solving for $\cos{\theta_o}$ and using the identity \eqref{eq:JacobiSNDN} yields
\begin{align}
	\label{eq:InversionB}
	\frac{\cos{\theta_o}}{\sqrt{u_-}}=h\dn\pa{\sqrt{u_-a^2}\pa{\tau+\nu_\theta\mathcal{G}_\theta^s}\left|1-\frac{u_+}{u_-}\right.},
\end{align}
where in taking the square root, we chose the branch $h=\pm1$ to obtain the motion in the correct hemisphere via Eq.~\eqref{eq:h}.  Although Eq.~\eqref{eq:InversionB} was derived under the assumption that a turning point has not yet been reached, it in fact continues properly past turning points to provide the full parameterized trajectory $\theta_o(\tau)$, as before.  Noting that when $k<0$, $\dn(\varphi|k)$ oscillates smoothly between $+1$ and $+\sqrt{1-k}$ with period $2K(k)$, we see that Eq.~\eqref{eq:InversionB} defines a trajectory $\theta_o(\tau)$ that oscillates between $\theta_-$ and $\theta_+$ with half-period $\hat{G}_\theta$.  Thus, it has the correct quantitative behavior at turning points, and we need only check that it satisfies the squared differential equation \eqref{eq:AngularODE} is satisfied, which is easily verified using the elliptic identities \eqref{eq:JacobiSNCN} and \eqref{eq:JacobiSNDN}.  This completes the proof that Eq.~\eqref{eq:InversionB} is the unique solution for $\theta_o(\tau)$ with initial conditions $\theta_o(0)=\theta_s$ and $\sign\br{\theta_o'(0)}=\nu_\theta$.

\subsubsection{Path integrals as functions of Mino time}

We have $G_\theta=\tau$ by definition, and the other path integrals may be expressed in terms of Mino time $\tau$ by the same method as in the ordinary case above.  Before a turning point is reached, we once again have Eqs.~\eqref{eq:DefiniteGphi} and \eqref{eq:DefiniteGt}.  Applying the formula \eqref{eq:JacobiAmplitude} to Eq.~\eqref{eq:AlmostInversionB} yields
\begin{align}
	\label{eq:UnfoldingB}
	\arcsin{\sqrt{\frac{\cos^2{\theta_o}-u_-}{u_+-u_-}}}=-h\nu_\theta\Upsilon_\tau,
\end{align}
where the (monotonically increasing in $\tau$) amplitude is
\begin{align}
	\Upsilon_\tau=\am\pa{\sqrt{u_-a^2}\pa{\tau+\nu_\theta\mathcal{G}_\theta^s}\left|1-\frac{u_+}{u_-}\right.},
\end{align}
and the restriction $|\varphi|\le K(k)$ is satisfied on account of our assumption that a turning point has not yet been reached.  Plugging Eq.~\eqref{eq:UnfoldingB} into Eqs.~\eqref{eq:DefiniteGphi} and \eqref{eq:DefiniteGt} as needed, and recalling that both $\Pi(n;-\varphi|k)=-\Pi(n;\varphi|k)$ and $E'(-\varphi|k)=-E'(\varphi|k)$ are odd in $\varphi$, we then find

\begin{align}
	G_\phi&=\frac{1}{\pa{1-u_-}\sqrt{u_-a^2}}\Pi\pa{\frac{u_+-u_-}{1-u_-};\Upsilon_\tau\left|1-\frac{u_+}{u_-}\right.}\nonumber\\
	\label{eq:GphiIntegralB}
	&\quad-\nu_\theta\mathcal{G}_\phi^s,\\
	\label{eq:GtIntegralB}
	G_t&=\sqrt{\frac{u_-}{a^2}}E\pa{\Upsilon_\tau\left|1-\frac{u_+}{u_-}\right.}-\nu_\theta\mathcal{G}_t^s.
\end{align}
Although Eqs.~\eqref{eq:GphiIntegralB} and \eqref{eq:GtIntegralB} were derived under the assumption that a turning point does not occur, they in fact extend properly through turning points to give the complete path integrals $G_\phi$ and $G_t$, as follows.  Since $\am(\varphi|k)$, $\Pi(n;\varphi|k)$ and $E(\varphi|k)$ are real and smooth functions of $\varphi$ provided $\max(k,n)<1$ [satisfied here for $k=1-u_+/u_-$ and $n=(u_+-u_-)/(1-u_-)$], it follows that the candidate formulas for $G_\phi$ and $G_t$ are real and smooth.  Thus, we need only check the differential equations \eqref{eq:GphiODE} and \eqref{eq:GtODE}, which is straightforward using the identity \eqref{eq:JacobiSNDN}.  This completes the proof that Eqs.~\eqref{eq:GphiIntegralB} and \eqref{eq:GtIntegralB} give the full path integrals \eqref{eq:Gphi} and \eqref{eq:Gt}.

Notice that Eq.~\eqref{eq:InversionB} may also be put in a similar form using $\sn(\varphi|k)=\sin\pa{\am(\varphi|k)}$ together with Eq.~\eqref{eq:JacobiSNDN},
\begin{align}
	\cos{\theta_o(\tau)}=h\sqrt{u_-+\pa{u_+-u_-}\sin^2{\Upsilon_\tau}}.
\end{align}

\subsubsection{Path integrals in terms of turning points}

Finally, it is useful for some purposes to express the path integrals $G_i$ as functions of $\theta_s$, $\theta_o$, $\nu_\theta$, and the number of turning points $m$ encountered along the trajectory.  A general treatment is given in App.~\ref{app:UnpackingPathIntegrals} below.  Therein, it was assumed that $x_-<x_+$, so we must take $x_\pm=\theta_\pm$ when $h=+1$ and $x_\pm=\theta_\mp=$ when $h=-1$; that is, $x_\pm=\theta_{\pm h}$.  Similarly, we have $\mathcal{H}_\pm=\mathcal{G}_\pm$ when $h=+1$ and $\mathcal{H}_\pm=\mathcal{G}_\mp$ when $h=-1$; that is,  $\mathcal{H}_\pm=\mathcal{G}_{\pm h}$.  Our choice of antiderivatives \eqref{eq:GthetaAntiderivativeB}, \eqref{eq:GphiAntiderivativeB}, and \eqref{eq:GtAntiderivativeB} have the property that $\mathcal{G}_i^+$ vanish for both $h=\pm1$.  Taking these facts into account, Eq.~\eqref{eq:EvenOddH} reduces to
\begin{align}
	G_i&=
	\begin{cases} 
		m\hat{G}_i+\nu_\theta\pa{\mathcal{G}_i^o-\mathcal{G}_i^s}
		&m\text{ even},\\
		\pa{m-h\nu_\theta}\hat{G}_i-\nu_\theta\pa{\mathcal{G}_i^o+\mathcal{G}_i^s}
		&m\text{ odd},
	\end{cases}\\
	&=\br{m-h\nu_\theta\frac{1-(-1)^m}{2}}\hat{G}_i+\nu_\theta\br{(-1)^m\mathcal{G}_i^o-\mathcal{G}_i^s},\nonumber
\end{align}
for $i\in\cu{\theta,\phi,t}$, in agreement with Eqs.~(81) of Ref.~\cite{Kapec2019} (choosing the upper sign therein).

\subsection{Unified inversion formula}

In the ordinary case $\eta>0$, we gave antiderivatives \eqref{eq:GthetaAntiderivativeA}, \eqref{eq:GphiAntiderivativeA}, and \eqref{eq:GtAntiderivativeA} involving elliptic integrals with parameter $u_+/u_-$, whereas in the vortical case $\eta<0$, we gave antiderivatives \eqref{eq:GthetaAntiderivativeB}, \eqref{eq:GphiAntiderivativeB}, and \eqref{eq:GtAntiderivativeB} involving elliptic integrals with parameter $1-u_+/u_-$.

Since $u_+>0$ and $\sign\pa{u_-}=-\sign\pa{\eta}$, these choices ensure that the parameter of any elliptic integral is always negative, so that the antiderivatives are real and smooth over the relevant domain.  On the other hand, the parameters exceed unity outside their domain, in which case the elliptic integrals becomes complex and suffer branch cut discontinuities.  Thus, while an antiderivative in one case is also an antiderivative for the other, it is not $C^1$ and cannot (in general) be used to compute definite integrals.

However, the manipulations carried out to check that the inversion formulas \eqref{eq:InversionA} and \eqref{eq:InversionB} satisfy the squared differential equation \eqref{eq:AngularODE} did not depend on any assumptions about the sign of $\eta$.  That is, each of the inversion formulas obeys the correct differential equation in both cases.  If these formulas also obey the correct initial conditions $\theta_o(0)=\theta_s$ and $\sign\pa{\theta_o'(0)}=\nu_\theta$ in both cases, then we conclude that they in fact remain valid in both cases.  Checking explicitly, we find that the initial value is correct, while the initial sign of derivative is incorrect.  However, this is easily adjusted by a simple sign flip, giving a unified inversion formula,
\begin{align}
	\label{eq:UnifiedInverseTheta}
	\cos{\theta_o(\tau)}&=-\sign\pa{\eta}\nu_\theta\sqrt{u_+}\sin{\Psi_\tau},
\end{align}
where Eq.~\eqref{eq:InversionAmplitudeA} for $\Psi_\tau$ is extended to the vortical case by 
\begin{align}
	\Psi_\tau=\am\pa{\sqrt{-u_-a^2}\br{\tau+\sign\pa{\eta}\nu_\theta\mathcal{G}_\theta^s}\left|\frac{u_+}{u_-}\right.},
\end{align}
using Eq.~\eqref{eq:GthetaAntiderivativeA} for $\mathcal{G}_\theta^s$.  Regardless of the sign of $\eta$, Eq.~\eqref{eq:UnifiedInverseTheta} satisfies the differential equation \eqref{eq:AngularODE} with initial conditions $\theta_o(0)=\theta_s$ and $\sign\br{\theta_o'(0)}=\nu_\theta$, and hence is a correct formula for the full parameterized trajectory $\theta_o(\tau)$.  This equivalence was first derived globally in Ref.~\cite{Kapec2019} using elliptic identities.

The expressions for $\theta_o(\tau)$ correctly extend outside their domain because they only involve $\sn(\varphi|k)$, which is a meromorphic complex function.  On the other hand, this equivalence breaks down for $G_\phi$ and $G_t$, as they involve elliptic integrals with branch cuts in the complex plane.

\section{Radial potential}
\label{sec:RadialPotential}

We now turn to the analysis of the radial potential $\mathcal{R}(r)$, which may be expressed as
\begin{align}
	\label{eq:RadialPotentialEta}
	\mathcal{R}(r)=\pa{r^2+a^2-a\lambda}^2-\zeta\Delta(r),
\end{align}
with
\begin{align}
	\label{eq:zeta}
	\zeta=\eta+\pa{\lambda-a}^2
	\geq0.
\end{align}
The restriction $\zeta\geq0$ follows from the constraints \eqref{eq:Protrusion} on the range of $\eta$.  The trajectories $\zeta=0$ saturating the bound \eqref{eq:zeta} are the principal null congruences \cite{Chandrasekhar1983}, with conserved quantities
\begin{align}
	\pa{\lambda,\eta}=\pa{a\sin^2{\theta_0},-a^2\cos^4{\theta_0}}.
\end{align}
Since principal null geodesics have $u_\pm=\cos^2{\theta_0}$, they stay at fixed $\theta=\theta_0$, where both the angular potential \eqref{eq:AngularPotential} and its derivative vanish, $\Theta(\theta_0)=\Theta'(\theta_0)=0$.  In this case, the roots of the radial potential \eqref{eq:RadialPotentialEta} are simply
\begin{align}
	\label{eq:RootsPNC}
	r=\pm ia\cos{\theta_0}.
\end{align}
For the remainder of this section, we assume that $\zeta\neq0$, in which case the bound \eqref{eq:zeta} becomes strict,
\begin{align}
	\label{eq:BoundPNC}
	\zeta>0.
\end{align}
We now find and classify the roots of the quartic radial potential \eqref{eq:RadialPotentialEta}.  We use Ferrari's method to express the four roots in a convenient form, and then study the special cases in which one or more roots coincide.  These cases define the boundaries between regions of the $(\lambda,\eta)$ parameter space in which the roots display different qualitative behaviors.

\subsection{Calculation of roots}

We now solve for the roots of the radial potential \eqref{eq:RadialPotentialEta} in the allowed range \eqref{eq:Protrusion}.  The analysis of a quartic polynomial usually begins by performing a simple scaling and translation to bring it into depressed form,
\begin{align}
	\label{eq:DepressedQuartic}
	r^4+\mathcal{A}r^2+\mathcal{B}r+\mathcal{C}=0,
\end{align}
which our potential \eqref{eq:RadialPotentialEta} already takes, with coefficients
\begin{align}
	\mathcal{A}&=a^2-\eta-\lambda^2, \\
	\mathcal{B}&=2M\br{\eta+\pa{\lambda-a}^2} >0, \\
	\mathcal{C}&=-a^2\eta.
\end{align}
Here, the positivity of $\mathcal{B}$ follows from Eq.~\eqref{eq:BoundPNC}.  Ferrari's method gives the general solution of the quartic \eqref{eq:DepressedQuartic} as
\begin{align}
	\label{eq:Ferrari}
	r=\frac{\pm_1\sqrt{2\xi_0}\pm_2\sqrt{-\pa{2\mathcal{A}+2\xi_0\pm_1\frac{\sqrt{2}\mathcal{B}}{\sqrt{\xi_0}}}}}{2},
\end{align}
where the four choices of sign $\pm_{1,2}$ yield the four roots, and where $\xi_0$ is any root of the ``resolvent cubic''
\begin{align}
	\label{eq:ResolventCubic}
	\mathsf{R}(\xi)=\xi^3+\mathcal{A}\xi^2+\pa{\frac{\mathcal{A}^2}{4}-\mathcal{C}}\xi-\frac{\mathcal{B}^2}{8}.
\end{align}
To obtain such a root, we first let $\xi=t-\mathcal{A}/3$ to bring the resolvent cubic into the depressed form
\begin{align}
	\mathsf{R}(t)=t^3+\mathcal{P}t+\mathcal{Q},
\end{align}
with coefficients
\begin{align}
	\mathcal{P}&=-\frac{\mathcal{A}^2}{12}-\mathcal{C}, \\
	\mathcal{Q}&=-\frac{\mathcal{A}}{3}\br{\pa{\frac{\mathcal{A}}{6}}^2-\mathcal{C}}-\frac{\mathcal{B}^2}{8}.
\end{align}
Cardano's method then gives 
\begin{align}
	\xi_0&=\omega_++\omega_--\frac{\mathcal{A}}{3},\\
	\xi_1&=e^{2\pi i/3}\omega_++e^{-2\pi i/3}\omega_--\frac{\mathcal{A}}{3},\\
	\xi_2&=e^{-2\pi i/3}\omega_++e^{2\pi i/3}\omega_--\frac{\mathcal{A}}{3},
\end{align}
as the roots of the original cubic \eqref{eq:ResolventCubic}, with
\begin{align}
	\label{eq:Cardano}
	\omega_\pm&=\sqrt[3]{-\frac{\mathcal{Q}}{2}\pm\sqrt{\pa{\frac{\mathcal{P}}{3}}^3+\pa{\frac{\mathcal{Q}}{2}}^2}}\\
	&=\sqrt[3]{-\frac{\mathcal{Q}}{2}\pm\sqrt{-\frac{\triangle_3}{108}}}.
\end{align}
In the last step, we introduced the discriminant $\triangle_3$ of the depressed cubic,
\begin{align}
	\triangle_3=-2^2\mathcal{P}^3-3^3\mathcal{Q}^2.
\end{align}
In Eq.~\eqref{eq:Cardano}, $\sqrt[3]{x}$ denotes either the real cube root of $x$, if $x$ is real, or else, the principal value of the cube root function (that is, the cubic root with maximal real part).

We have chosen this method for solving the cubic because it guarantees that $\xi_0$ is always real and positive,
\begin{align}
	\xi_0>0.
\end{align}
To see this, consider separately the cases where $\triangle_3<0$ and $\triangle_3>0$.  If $\triangle_3<0$, then $\omega_\pm$ are real, implying that $\xi_0$ is real and $\xi_1=\bar{\xi}_2$ are complex conjugates.  If $\triangle_3>0$, then $\omega_+=\bar{\omega}_-$ are complex conjugates, implying that all three roots $\xi_0$, $\xi_1$ and $\xi_2$ are real.  In that case, $\xi_0$ is the largest root, since by the definition of $\sqrt[3]{x}$, $\omega_+$ has a larger real part than either of the other two cube roots of $\omega_+^3$ (and likewise for $\omega_-$); that is, we have $\textrm{Re}[\omega_\pm]>\textrm{Re}[e^{2\pi i/3}\omega_\pm]$ and $\textrm{Re}[\omega_\pm]>\textrm{Re}[e^{-2\pi i/3}\omega_\pm]$.  Thus, in all cases, $\xi_0$ is the largest real root.  Finally, since the original polynomial \eqref{eq:ResolventCubic} ranges from $\mathsf{R}(0)=-\mathcal{B}^2/8<0$ to $\mathsf{R}(+\infty)=+\infty$ over the positive real axis, it always admits at least one positive real root.  This proves that $\xi_0$ is always real and positive (and the largest such root).

We now use this root $\xi_0$ in Ferrari's formula \eqref{eq:Ferrari} for the solution of the quartic.  Defining
\begin{align}
	\label{eq:z}
	z=\sqrt{\frac{\xi_0}{2}}>0,
\end{align}
the four roots are obtained in the particularly simple form
\begin{subequations}
\label{eq:RadialRoots}
\begin{align}
	r_1&=-z-\sqrt{-\frac{\mathcal{A}}{2}-z^2+\frac{\mathcal{B}}{4z}}, \\
	r_2&=-z+\sqrt{-\frac{\mathcal{A}}{2}-z^2+\frac{\mathcal{B}}{4z}},\\
	r_3&=z-\sqrt{-\frac{\mathcal{A}}{2}-z^2-\frac{\mathcal{B}}{4z}}, \\
	r_4&=z+\sqrt{-\frac{\mathcal{A}}{2}-z^2-\frac{\mathcal{B}}{4z}}.
\end{align}
\end{subequations}
These names are natural because of the ordering of the roots, discussed in Sec.~\ref{sec:Classification} below.  Notice that
\begin{align}
	\label{eq:Vieta}
	r_1+r_2+r_3+r_4=0.
\end{align}
This vanishing of the roots' sum is a general property of depressed quartics (generalized by Vieta's formulas).

In the special case of extreme Kerr ($a=M$), the roots can also be expressed in the simpler form \cite{Gates2018}
\begin{align}
	\label{eq:ExtremeRoots}
	r&=\pm_1\triangle_r\pm_2\sqrt{\pa{\triangle_r\mp_1M}^2+M\pa{\lambda-2M}},\\
	\triangle_r&=\frac{1}{2}\sqrt{\eta+\pa{\lambda-M}^2}
	=\sqrt{\frac{\mathcal{B}}{8M}},
\end{align}
where the two independent sign choices $\pm_1$ and $\pm_2$ give the four roots, but without a clear ordering.

\subsection{Classification of roots}
\label{sec:Classification}

We now determine the character (real or complex) and ordering of the roots $\cu{r_1,r_2,r_3,r_4}$ as a function of the conserved quantities $\lambda$ and $\eta$.  The boundaries between regions of different behaviors correspond to conserved quantities $(\lambda,\eta)$ such that one or more roots coincide, so we begin by classifying these critical cases.

The maximally degenerate case of all four roots coinciding occurs when the quartic is just proportional to $r^4$, i.e., when
\begin{align}
	\check{r}=0,\quad
	\check{\lambda}=a,\quad
	\check{\eta}=0.
\end{align}
This is the equatorial principal null geodesic, with $\zeta=0$ and $\theta_0=\pi/2$ [see Eq.~\eqref{eq:RootsPNC}].

Triple roots occur when $\mathcal{R}''(\hat{r})=\mathcal{R}'(\hat{r})=\mathcal{R}(\hat{r})=0$, which straightforwardly implies
\begin{align}
	\label{eq:rhat}
	\hat{r}&=M\br{1-\pa{1-\frac{a^2}{M^2}}^{1/3}}\in\pa{0,r_-},\\
	\hat{\lambda}&=a-\frac{\hat{r}^2\pa{2\hat{r}-3M}}{Ma},\\
	\hat{\eta}&=6\hat{r}^2-\hat{\lambda}^2+a^2,
\end{align}
with $\hat{\zeta}=4\hat{r}^3/M>0$ satisfying Eq.~\eqref{eq:BoundPNC}.

\begin{figure*}
	\centering
	\includegraphics[width=\textwidth]{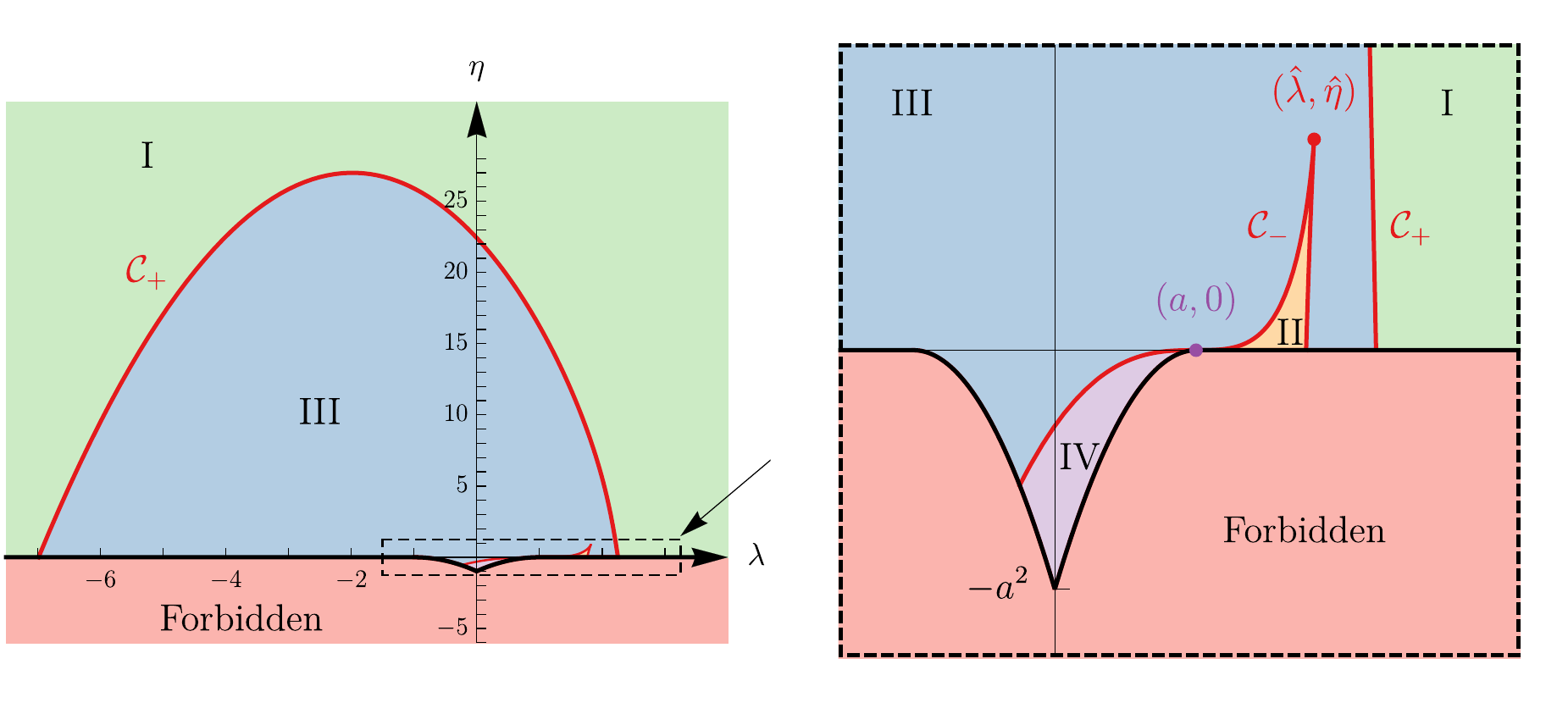}
	\caption{Regions of conserved quantity space corresponding to different qualitative behaviors of the roots of the radial potential.  Here we show the case of spin $a/M=99\%$ and set $M=1$.  The triple root $\hat{r}$ is shown with a red dot and the quadruple root $\check{r}$ with a purple dot.  In the low-spin limit $a\to0$, regions II and IV disappear, while in the high-spin limit $a \to M$ the rightmost portion of $\mathcal{C}_-$ merges with $\mathcal{C}_+$, so that region II is adjacent to region I.}
	\label{fig:DetailedRegions}
\end{figure*}

Double roots occur when $\mathcal{R}(\tilde{r})=\mathcal{R}'(\tilde{r})=0$, i.e., when
\begin{align}
	\label{eq:FirstCondition}
	0&=\pa{\tilde{r}^2+a^2-a\tilde{\lambda}}^2-\zeta\Delta(\tilde{r}),\\
	\label{eq:SecondCondition}
	0&=4\tilde{r}\pa{\tilde{r}^2+a^2-a\tilde{\lambda}}-2\zeta\pa{\tilde{r}-M}.
\end{align}
If $\tilde{r}=M$, then these equations are satisfied if and only if $a=M$ and $\lambda=2M$, corresponding to the superradiant bound of an extreme black hole.  This is a very interesting regime that we exclude for present purposes, where we consider $0<a<M$.  Hence, we must have $\tilde{r}\neq M$, in which case  Eq.~\eqref{eq:SecondCondition} can be solved to find
\begin{align}
	\zeta=\frac{2\tilde{r}}{\tilde{r}-M}\pa{\tilde{r}^2+a^2-a\tilde{\lambda}}.
\end{align}
Next, plugging back into the first condition \eqref{eq:FirstCondition} yields
\begin{align}
	\pa{\tilde{r}^2+a^2-a\lambda}\br{\tilde{r}^2+a^2-a\tilde{\lambda}-\frac{2\tilde{r}\Delta(\tilde{r})}{\tilde{r}-M}}=0.
\end{align}
Since $\zeta\neq0$ by assumption, the first term is not allowed to vanish. Therefore, we are left with
\begin{align}
	\label{eq:CriticalAngularMomentum}
	\tilde{\lambda}=a+\frac{\tilde{r}}{a}\br{\tilde{r}-\frac{2\Delta(\tilde{r})}{\tilde{r}-M}}.
\end{align}
Back-substituting into Eq.~\eqref{eq:SecondCondition} then gives
\begin{align}
	\label{eq:CriticalCarterIntegral}
	\tilde{\eta}&=\frac{\tilde{r}^3}{a^2\pa{\tilde{r}-M}^2}\br{4Ma^2-\tilde{r}\pa{\tilde{r}-3M}^2}\\
	&=\frac{\tilde{r}^3}{a^2}\br{\frac{4M\Delta(\tilde{r})}{\pa{\tilde{r}-M}^2}-\tilde{r}},
\end{align}
with $\tilde{\zeta}=4\tilde{r}^2\pa{\tilde{r}-M}^{-2}\Delta(\tilde{r})>0$ satisfying Eq.~\eqref{eq:BoundPNC}.  The formulas \eqref{eq:CriticalAngularMomentum} and \eqref{eq:CriticalCarterIntegral} describe a curve in the $(\lambda,\eta)$-space parameterized by the radius $\tilde{r}$.  We now determine the portion of this curve within the allowed region \eqref{eq:Protrusion}.  Its edges occur at $\tilde{\eta}\in\{0,-(\tilde{\lambda}\pm a)^2\}$ depending on $\tilde{\lambda}$:
\begin{align}
	\label{eq:Edge1}
	\tilde{\eta}=0:&\quad
	4Ma^2-\tilde{r}\pa{\tilde{r}-3M}^2=0,\\
	\label{eq:Edge2}
	\tilde{\eta}=-(\tilde{\lambda}+a)^2:&\quad
	Ma^2+\tilde{r}^2\pa{2\tilde{r}-3M}=0,\\
	\tilde{\eta}=-(\tilde{\lambda}-a)^2:&\quad
	\tilde{r}=r_\pm
\end{align}
Note $\tilde{r}=0$ is also valid for $\tilde{\eta}=0$ and $\tilde{\eta}=-(\tilde{\lambda}-a)^2$.  The roots of the above cubic polynomials are all real and can thus be written using the trigonometric formulas
\begin{align}
	\label{eq:EdgeRoot1}
	\tilde{r}&=2M+2M\cos\br{\frac{2\pi k}{3}+\frac{2}{3}\arccos\pa{\frac{a}{M}}},\\
	\label{eq:EdgeRoot2}
	\tilde{r}&=\frac{M}{2}+M\cos\br{\frac{2\pi k}{3}+\frac{2}{3}\arcsin\pa{\frac{a}{M}}},
\end{align}
for $k\in\cu{0,1,2}$.  Eqs.~\eqref{eq:EdgeRoot1} and \eqref{eq:EdgeRoot2} are the solutions to the cubic equations in Eqs.~\eqref{eq:Edge1} and \eqref{eq:Edge2}, respectively.  

Eqs.~\eqref{eq:EdgeRoot1} and \eqref{eq:EdgeRoot2} together with $\tilde{r}=0$ and $\tilde{r}=r_\pm$ are the complete list of radii where a curve of double roots may intersect the edge of the allowed region \eqref{eq:Protrusion}.  Examining each case, we find that Eq.~\eqref{eq:Protrusion} is satisfied only in the ranges
\begin{align}
	\label{eq:CriticalRange1}
	\tilde{r}&\in\br{\tilde{r}_2,\tilde{r}_3}\quad
	\text{(outside horizon, defines $\mathcal{C}_+$)},\\
	\label{eq:CriticalRange2}
	\tilde{r}&\in\br{\tilde{r}_u,\tilde{r}_1}\quad
	\text{(inside horizon, defines $\mathcal{C}_-$)},
\end{align}
where we introduced special notation for the four relevant roots from Eqs.~\eqref{eq:EdgeRoot1} and \eqref{eq:EdgeRoot2},
\begin{align}
	\tilde{r}_u&=\frac{M}{2}+M\cos\br{\frac{2\pi}{3}+\frac{2}{3}\arcsin\pa{\frac{a}{M}}},\\
	\tilde{r}_1&=2M+2M\cos\br{\frac{2\pi}{3}+\frac{2}{3}\arccos\pa{\frac{a}{M}}},\\
	\tilde{r}_2&=2M+2M\cos\br{\frac{4\pi}{3}+\frac{2}{3}\arccos\pa{\frac{a}{M}}},\\
	\tilde{r}_3&=2M+2M\cos\br{\frac{2}{3}\arccos\pa{\frac{a}{M}}}.
\end{align}
These real roots have the ordering
\begin{align}
	\tilde{r}_u<0<\hat{r}<\tilde{r}_1<r_-<r_+<\tilde{r}_2<\tilde{r}_3.
\end{align}
The ranges \eqref{eq:CriticalRange1} and \eqref{eq:CriticalRange2} define two disjoint curves $\mathcal{C}_\pm$ in $(\lambda,\eta)$-space via $\lambda=\tilde{\lambda}(\tilde{r})$ and $\eta=\tilde{\eta}(\tilde{r})$ [see Fig.~\ref{fig:DetailedRegions}].  Note that the range \eqref{eq:CriticalRange1} of $\mathcal{C}_+$ is equivalent to $\tilde{\eta}\geq0$, so the orbits bound at double roots $\tilde{r}$ outside the horizon all cross the equatorial plane, with the boundary values $\tilde{r}_2$ and $\tilde{r}_3$ corresponding to prograde and retrograde circular equatorial ($\tilde{\eta}=0$) orbits, respectively. 

Instead of using $\tilde{r}$ as a parameter, we can instead express the curves as $\tilde{\eta}(\tilde{\lambda})$ using Eq.~\eqref{eq:CriticalCarterIntegral} and the inversion of the cubic equation \eqref{eq:CriticalAngularMomentum}.  Using the trigonometric cubic formula, we write the roots as
\begin{align}
	\tilde{r}^{(k)}(\tilde{\lambda})&=M+2M\triangle_{\tilde{\lambda}}\cos\br{\frac{2\pi k}{3}+\frac{1}{3}\arccos\pa{\frac{1-\frac{a^2}{M^2}}{\triangle_{\tilde{\lambda}}^3}}},\nonumber\\
	&\triangle_{\tilde{\lambda}}=\sqrt{1-\frac{a(a+\tilde{\lambda})}{3M^2}},
\end{align}
with $k\in\cu{0,1,2}$.  For spins $a/M\le1/\sqrt{2}$, the relevant inversions are then
\begin{align}
	\mathcal{C}_+:&\quad
	\tilde{r}=\tilde{r}^{(0)}(\tilde{\lambda})
	\in\br{\tilde{r}_2,\tilde{r}_3},\quad
	\tilde{\lambda}(\tilde{r}_3)\le\tilde{\lambda}\le\tilde{\lambda}(\tilde{r}_2),\\
	\label{eq:C-}
	\mathcal{C}_-:&\quad
	\tilde{r}=
	\begin{cases}
		\tilde{r}^{(2)}(\tilde{\lambda})
		\in\br{\hat{r},\tilde{r}_1},&
		\tilde{\lambda}(\tilde{r}_1)\le\tilde{\lambda}\le\tilde{\lambda}(\hat{r}),\\
		\tilde{r}^{(1)}(\tilde{\lambda})
		\in\br{\tilde{r}_u,\hat{r}},&
		\tilde{\lambda}(\hat{r})\le\tilde{\lambda}\le\tilde{\lambda}(\tilde{r}_u),
	\end{cases}
\end{align}
where $\hat{r}$ denotes the triple root of the radial potential, given in Eq.~\eqref{eq:rhat} above.  When $a/M>1/\sqrt{2}$, the photon shell intersects the ergosphere, which extends up to equatorial radius $2M>\tilde{r}_2$.  In that case, $\mathcal{C}_-$ retains the same description \eqref{eq:C-}, while $\mathcal{C}_+$ requires two segments,
\begin{align}
	\mathcal{C}_+:&\quad
	\tilde{r}=
	\begin{cases}
		\tilde{r}^{(0)}(\tilde{\lambda})
		\in\br{2M,\tilde{r}_3},&
		\tilde{\lambda}(\tilde{r}_3)\le\tilde{\lambda}\le\tilde{\lambda}_e,\\
		\tilde{r}^{(2)}(\tilde{\lambda})
		\in\br{\tilde{r}_2,2M},&
		\tilde{\lambda}_e\le\tilde{\lambda}\le\tilde{\lambda}(\tilde{r}_2),
	\end{cases}
\end{align}
where $\tilde{\lambda}_e:=\tilde{\lambda}(2M)=-a+3M^2/a$ denotes the angular momentum of the spherical orbit bound at $\tilde{r}=2M$.  Plugging these expressions into Eq.~\eqref{eq:CriticalCarterIntegral} for $\tilde{\eta}$ gives each segment of the curve as a function $\tilde{\eta}(\tilde{\lambda})$.  The curve $\mathcal{C}_-$ has an inflection point at $\tilde{r}=0$ and a kink at $\tilde{r}=\hat{r}$ [see right panel in Fig.~\ref{fig:DetailedRegions}].

\begin{figure*}
	\includegraphics[width=\textwidth]{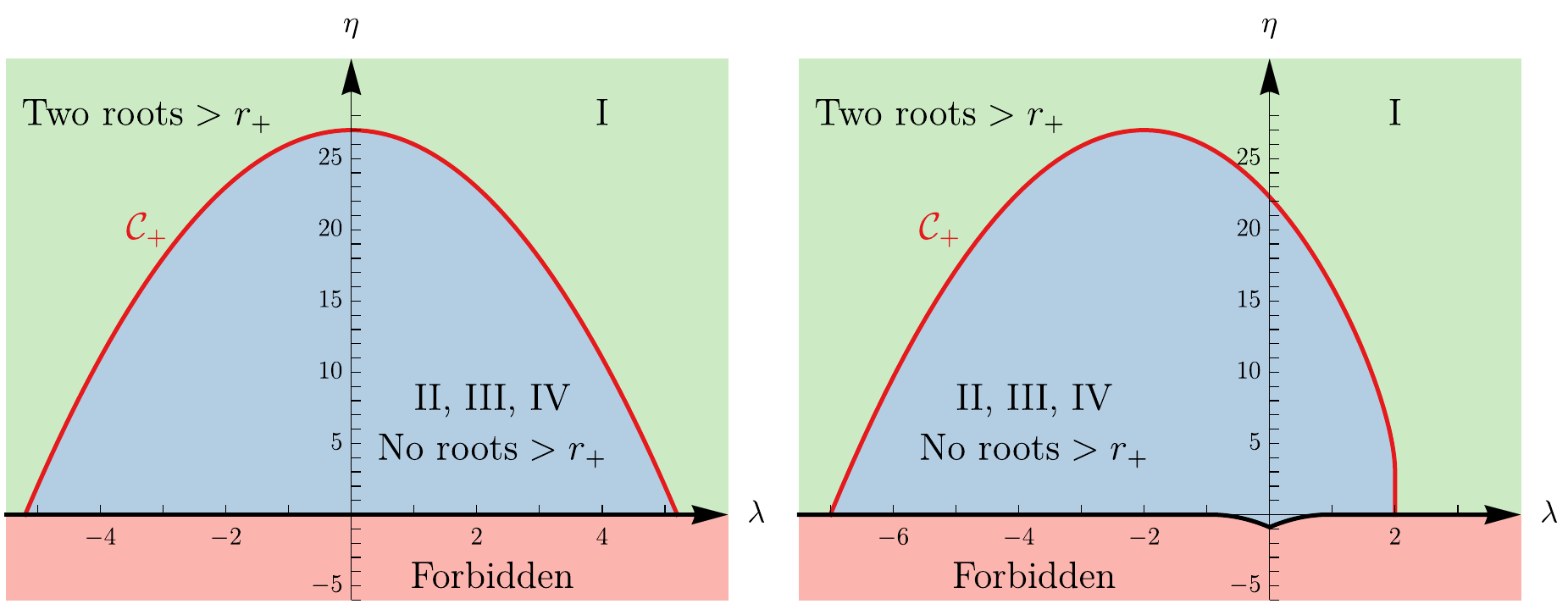}
	\caption{The outer critical curve $\mathcal{C}_+$ is the boundary between rays with two roots outside the horizon (region I) and with no roots outside the horizon (regions II, III, IV).  As the spin is increased from $a=0$ on the left to $a=M$ on the right, the region of vortical geodesics (lower protrusion) grows in size, while the right side of $\mathcal{C}_+$ tucks in and develops a vertical segment.  (This segment maps to the ``NHEKline'' on the image of an observer  \cite{Bardeen1973,Gralla2017}.)  We have set $M=1$ in these plots.}
	\label{fig:RadialRegions}
\end{figure*}

The critical curves $\mathcal{C}_\pm$ divide the allowed region of parameter space into four subregions as depicted in Fig.~\ref{fig:DetailedRegions}.  Since complex roots appear in conjugate pairs, and all roots must vary smoothly in the $(\lambda,\eta)$-plane, each subregion corresponds to a definite number of real roots (either zero, two, or four).  Furthermore, the expressions \eqref{eq:RadialRoots} for the roots $\cu{r_1,r_2,r_3,r_4}$ are smooth functions in each subregion, so any real roots retain their ordering throughout a subregion.  Moreover, real roots may move through the inner and outer event horizons only via a double root at the horizon,\footnote{The radial potential is nonnegative at both horizons and positive at infinity: $\mathcal{R}(r_\pm)=\pa{r_\pm^2+a^2-a\lambda}^2\ge0$ and $\mathcal{R}(\pm\infty)=+\infty$.  As such, there must always be an even number of real roots in each of the ranges $r<r_-$, $r_-<r<r_+$, and $r>r_+$.} meaning that real roots also retain their ordering relative to the horizons within each subregion.  Thus, to determine the character of the roots throughout any subregion, it suffices to evaluate the formulas \eqref{eq:RadialRoots} at a single point therein.  Doing so results in the general classification:
\begin{itemize}
	\item[I.] Four real roots, two outside horizon:\\
	$r_1<r_2<r_-<r_+<r_3<r_4$.
	\item[II.] Four real roots, all inside horizon:\\
	$r_1<r_2<r_3<r_4<r_-<r_+$.
	\item[III.] Two real roots, both inside horizon:\\
	$r_1<r_2<r_-<r_+$ and $r_3=\bar{r}_4$.
	\item[IV.] No real roots:
	$r_1=\bar{r}_2$ and $r_3=\bar{r}_4$.
\end{itemize}
Notice that there are never any real roots between the inner and outer horizons $r_-$ and $r_+$.  On the critical curve $\mathcal{C}_+$, we have $r_3=r_4>r_+$, while on the portion of $\mathcal{C}_-$ in the upper-half plane ($\eta>0$), we have $r_3=r_4<r_-$, with $\hat{r}=r_2=r_3=r_4<r_-$ at the triple-root $(\hat{\lambda},\hat{\eta})$.  On the portion of $\mathcal{C}_-$ in the lower-half plane ($\eta<0$), we have $r_1=r_2<r_-$, with all four roots coinciding at the intersection of $\mathcal{C}_-$ with the horizontal axis $\eta=0$.

The allowed range(s) of $r$ for each of the four cases can be determined by checking where the radial potential is positive for a single choice of conserved quantities in each region.  Noting that the potential is always positive at $r\to\pm\infty$, the ranges are: $r<r_1$, $r_2<r<r_3$, and $r>r_4$ for cases I and II; $r<r_1$ and $r>r_2$ for case III; and any value of $r$ for case IV.  Restricting to motion outside the horizon, the relevant ranges are thus:
\begin{itemize}
	\item[] Ia. $r_+<r<r_3$ (white hole to black hole).
	\item[] Ib. $r_4<r<\infty$ (scattering).
	\item[] II, III, IV. $r_+<r<\infty$ (fly in or out).
\end{itemize}
In case Ia, the ray emerges from the white hole, reaches a turning point at $r=r_3$, and falls into the black hole.  In case Ib, the ray enters from infinity, reaches a turning point at $r=r_4$, and returns to infinity.  In cases II, III, and IV, the ray either starts from the white hole horizon and ends at infinity, or starts from infinity and ends at the black hole horizon.  Fig.~\ref{fig:RadialRegions} illustrates these regions.

\subsection{Radial integrals and inversion}
\label{sec:RadialIntegrals}

The above classification of the radial motions enables the radial integrals $I_r$, $I_\phi$, $I_t$, and $I_\sigma$ to be expressed in manifestly real elliptic form using standard transformations.  The needed transformations group themselves into yet another logically distinct set of cases, according to the turning point(s) of the maximally extended trajectory:\footnote{In Case IV, the trajectory has no turning point ($-\infty<r<\infty$), but the Legendre form of the antiderivative we give is smooth only over the range $-z<r<\infty$, which covers the exterior since $z>0$.}
\begin{itemize}
	\item[(1)] Case Ia: $r_2<r<r_3$.
	\item[(2)] Cases Ib and II: $r_4<r<\infty$.
	\item[(3)] Case III: $r_2<r<\infty$.
	\item[(4)] Case IV: $-z<r<\infty$.
\end{itemize}
For each case (1)-(4), we proceed as with the polar motion above: first, we find smooth real antiderivatives for each integral; next, we invert $I_r=\tau$ to find $r_o(\tau)$; and finally, we find expressions for $I_\phi$ and $I_t$, both as functions of $\tau$ and expressed in terms of the number of turning points.  Since the method is essentially the same as in the polar case (but lengthier), we defer treatment to App.~\ref{app:RadialIntegrals}.  The results are summarized in Sec.~\ref{sec:Summary} below.

\section{Comparison to Previous Work}
\label{sec:Comparison}

We now compare our results to previous work.  For the polar motion, the formulas for the roots, classification of motion types, reduction to elliptic integrals, and inversion for $\theta_o(\tau)$ have all appeared before in the literature.  We provide a unified presentation, introduce a method of derivation that generalizes to the radial case, and also give for the first time the Mino-time parameterization of the path integrals $G_\phi(\tau)$ and $G_t(\tau)$.  For the radial motion, the roots had not been explicitly solved for, the complete list of motion types had not been associated with regions of conserved quantity space (Fig.~\ref{fig:DetailedRegions}), only a subset of the integral reductions and inversions had been performed, and formulas for $I_\phi(\tau)$ and $I_t(\tau)$ had not previously appeared.

We now make a more detailed comparison to a subset of earlier work.  Rauch and Blandford \cite{Rauch1994} used the standard substitutions \cite{Byrd1954} (the same ones we use) to reduce $I_r$ to Legendre elliptic form in all possible cases.  Dexter and Agol \cite{Dexter2009} reduced $I_r$ and $G_\theta$ to Carlson symmetric form and found the inversion formulas $r(\tau)$ and $\theta(\tau)$ in a subset of cases.  Esteban \& V\'asquez \cite{Vazquez2004} expressed a subset of the path integrals explicitly in terms of the number of turning points, and Kapec and Lupsasca \cite{Kapec2019} obtained corresponding (and simplified) expressions for all of the angular path integrals.  We go beyond these works by delineating the regions of conserved quantity space where each case of radial motion applies, by finding explicit ordered expressions for the radial roots, by finding inversion formulas valid in all cases, and by computing all geodesic integrals $G_\theta$, $G_\phi$, $G_t$, $I_r$, $I_\phi$, $I_t$, and $I_\sigma$ in all cases.

Analytic solutions for $x^\mu(\tau)$ were given previously by Hackmann \cite{Hackmann2010} using the Weierstrass elliptic function.  These expressions are slightly less explicit than ours, since they require the computation of an integral to relate to given initial data, need manual gluing at some turning points, and also feature a reference root to be found in each case.  (Our explicit solution for the roots simplifies the latter task.)  The solutions of Ref.~\cite{Hackmann2010} also appear to be less general, since the reference root is assumed to be real, and it is not clear whether the results extend to case IV, where all roots are complex.  However, Hackmann's approach goes beyond our work in treating timelike geodesics as well.

Finally, we note that an approach similar to ours was followed in Refs.~\cite{Drasco2004,Fujita2009} to analyze bound timelike geodesics in Mino time.

\section{Recipe for trajectories}
\label{sec:Summary}

We now explain how to use the results of this paper to construct a parameterized trajectory for a given set of initial conditions, excluding certain measure-zero cases.  Beginning with the initial position $x_s^\mu$ and momentum $p_s^\mu$, one first determines $\lambda$ and $\eta$ via Eqs.~\eqref{eq:ConservedQuantities}--\eqref{eq:RescaledConservedQuantities}.  Next, one determines the type of polar motion (type A or B) according to whether $\eta$ is positive or negative, respectively.  One then evaluates the roots $\cu{r_1,r_2,r_3,r_4}$ [Eqs.~\eqref{eq:RadialRoots}] to determine the radial case I, II, III, or IV.  (One way to do so is the following: If $r_2$ is not real, the motion is case IV; if $r_2$ is real, then the motion is case III, II, or I if $r_4$ is complex, real but inside the horizon, or real and outside the horizon, respectively.)  Next, one determines the substitution class (1), (2), (3), or (4) according to the following: If case I, choose (1) or (2) according to whether the initial radius is less than $r_3$ or greater than $r_4$; if case II, III, or IV, choose (2), (3), or (4), respectively.  

Having determined the appropriate type of polar motion (A or B) and radial motion (1)-(4), the trajectories are given in a unified notation in the relevant subsections of the paper.  As an example, we will consider polar type A and radial type (2) (rays that enter and leave via the celestial sphere).  The solution is given in the notation $x_o^\mu(\tau)$, where $\nu_\theta$ and $\nu_r$ are the initial signs of the polar momentum $p_s^\theta$ and radial momentum $p_s^r$, respectively.  The polar motion $\theta_o(\tau)$ is given in Eq.~\eqref{eq:InversionA}.  The radial motion $r_o(\tau)$ is given in Eq.~\eqref{eq:Inversion2}.  The azimuthal motion $\phi_o(\tau)$ is given by Eqs.~\eqref{eq:Deltaphi} and \eqref{eq:IphiDecomposition} using Eqs.~\eqref{eq:RadialCase2} and \eqref{eq:GphiIntegralA}.  The temporal motion $t_o(\tau)$ is given by Eqs.~\eqref{eq:Deltat} and \eqref{eq:ItDecomposition} using Eqs.~\eqref{eq:RadialCase2} and \eqref{eq:GtIntegralA}.

\acknowledgements{SEG was supported by NSF grant PHY-1752809 to the University of Arizona.  This work was initiated at the Aspen Center for Physics, which is supported by National Science Foundation grant PHY-1607611.  AL was supported in part by the Jacob Goldfield Foundation.  We thank Torben Frost for identifying some errors in the original manuscript.
\hfill\includegraphics[scale=.01]{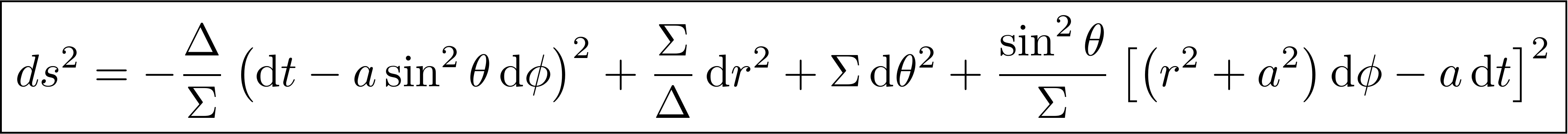}}

\appendix

\section{Unpacking path integrals}
\label{app:UnpackingPathIntegrals}

\begin{figure}[htp]
	\centering
	\includegraphics[width=.85\columnwidth]{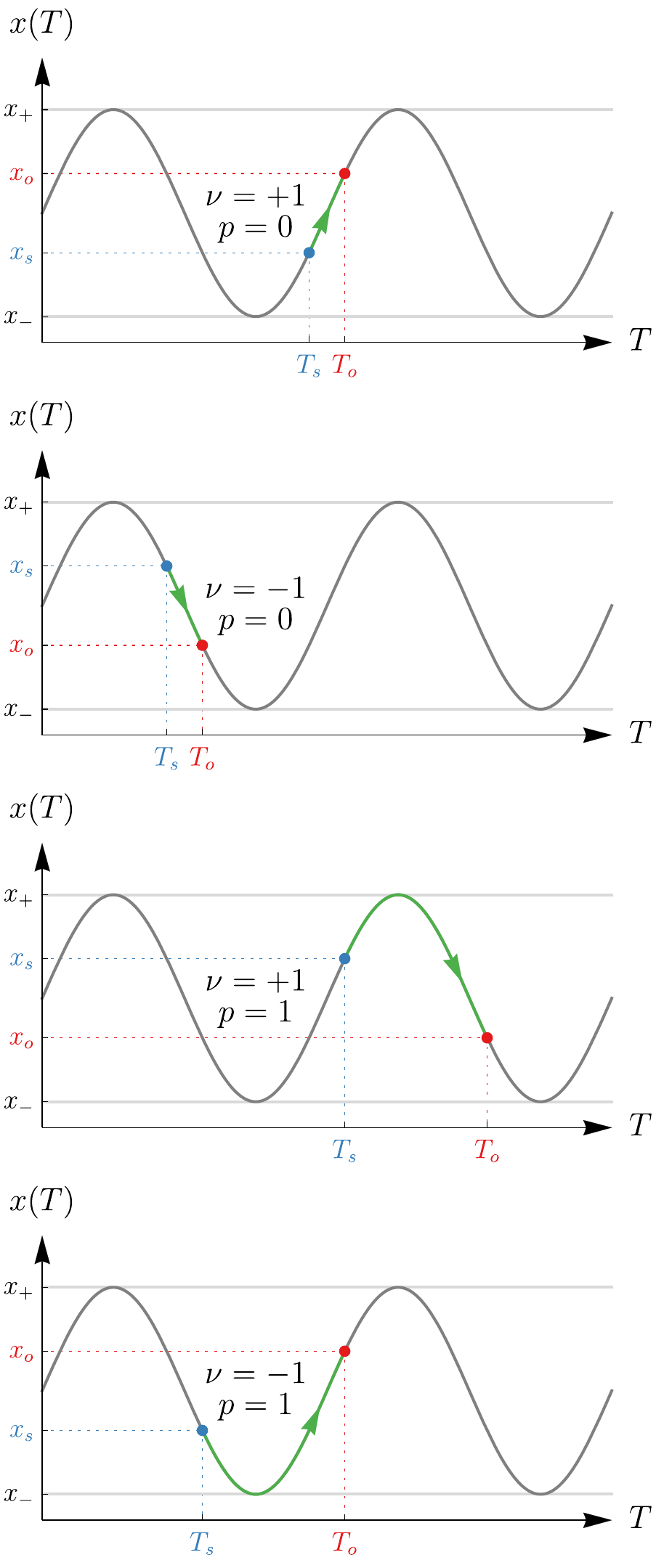}
	\caption{Illustration of the four cases considered in Eqs.~\eqref{eq:OscillationCases}.}
	\label{fig:AngularPathIntegral}
\end{figure}

Consider a ``trajectory'' $x(T)$ defined between $T_s$ and $T_o$ that periodically oscillates in some range $x\in\br{x_-,x_+}$.  Given a function $h(x)$ that is real and smooth in this range (and integrable as the edges are approached), we may define a path integral
\begin{align}
	H=\int_{T_s}^{T_o}h(x(T))\ab{\frac{dx}{dT}}\ed T.
\end{align}
Over any segment between adjacent turning points, the integrand is simply $\pm h(x)\ed x$, with $\pm$ the sign of $dx/dT$.  We thus abbreviate the integral using the slash notation
\begin{align}
	H=\fint_{x_s}^{x_o}\pm h(x)\ed x,
\end{align}
where $x(T_s)=x_s$ and $x(T_o)=x_o$, while the $\pm$ is always the sign of $dx/dT$ (and hence switches at turning points).  This makes it manifest that the integral depends only on the initial value $x_s$, the initial sign $\nu=\sign\pa{x'(T_s)}$, the final value $x_o$, and the number of turning points $p$ encountered along the way,
\begin{align}
	\label{eq:H}
	H=H(x_s,\nu,x_o,p).
\end{align}
We now find explicit expressions for $H$ in terms of these parameters, as well as any choice of (real and smooth) antiderivative $\mathcal{H}(x)$ such that
\begin{align}
	\frac{d\mathcal{H}}{dx}=h(x).
\end{align}
We will use the notation 
\begin{align}
	\mathcal{H}_i=\mathcal{H}(x=x_i),\quad
	i\in\cu{s,o,+,-}.
\end{align}
The path integral has the quasiperiodicity property
\begin{align}
	\label{eq:PeriodicityH}
	H(p+2)=H(p)+2\hat{H},
\end{align}
where the other arguments in Eq.~\eqref{eq:H} are fixed, while
\begin{align}
	\hat{H}=\int_{x_-}^{x_+}H(x)\ed x
	=\mathcal{H}_+-\mathcal{H}_-,
\end{align}
denotes the path integral over a half-libration.  Thus, it suffices to consider four cases: $\nu=\pm$ and $m\in\cu{0,1}$.  These are depicted in Fig.~\ref{fig:AngularPathIntegral}.  Going through each of these configurations in turn, we find
\begin{subequations}
\label{eq:OscillationCases}
\begin{align}
	\nu=+1,\ p=0:&\quad
	H=\mathcal{H}_o-\mathcal{H}_s,\\
	\nu=-1,\ p=0:&\quad
	H=\mathcal{H}_s-\mathcal{H}_o,\\
	\nu=+1,\ p=1:&\quad
	H=-\mathcal{H}_s-\mathcal{H}_o+2\mathcal{H}_+,\\
	\nu=-1,\ p=1:&\quad
	H=\mathcal{H}_s+\mathcal{H}_o-2\mathcal{H}_-.
\end{align}
\end{subequations}
Pairing these expressions by their value of $p$, and then promoting them to arbitrary integer $p$ using Eq.~\eqref{eq:PeriodicityH}, we finally obtain
\begin{align}
	\label{eq:EvenOddH}
	H=
	\begin{cases}
		p\hat{H}+\nu\pa{\mathcal{H}_o-\mathcal{H}_s}
		&p\text{ even},\\
		p \hat{H}+\nu\pa{\mathcal{H}_++\mathcal{H}_--\mathcal{H}_o-\mathcal{H}_s}
		&p\text{ odd},
	\end{cases}
\end{align}
which correctly reproduces Eqs.~\eqref{eq:OscillationCases} when $p\in\cu{0,1}$.  Finally, the even and odd cases can be combined into
\begin{align}
	\label{eq:CombinedH}
	H=p\hat{H}+\nu\br{(-1)^p\mathcal{H}_o-\mathcal{H}_s+\frac{1-(-1)^p}{2}\pa{\mathcal{H}_++\mathcal{H}_-}}.
\end{align}

\onecolumngrid

\section{Radial integrals and inversion}
\label{app:RadialIntegrals}

In this appendix, we analyze the radial integrals and trajectories following the approach established in the polar case in Sec.~\ref{sec:AngularPotential}.  It is convenient to rewrite the radial integrals as
\begin{align}
	I_r&=I_0,\\
	\label{eq:IphiDecomposition}
	I_\phi&=\frac{2Ma}{r_+-r_-}\br{\pa{r_+-\frac{a\lambda}{2M}}I_+-\pa{r_--\frac{a\lambda}{2M}}I_-},\\
	\label{eq:ItDecomposition}
	I_t&=\frac{(2M)^2}{r_+-r_-}\br{r_+\pa{r_+-\frac{a\lambda}{2M}}I_+-r_-\pa{r_--\frac{a\lambda}{2M}}I_-}+(2M)^2I_0+(2M)I_1+I_2,
\end{align}
where we introduced
\begin{align}
	\label{eq:EllipticIntegralsBasis}
	I_\pm&=\fint_{r_s}^{r_o}\frac{\ed r}{\pm_r\pa{r-r_\pm}\sqrt{\mathcal{R}(r)}},\qquad
	I_\ell=\fint_{r_s}^{r_o}\frac{r^\ell\ed r}{\pm_r\sqrt{\mathcal{R}(r)}},
\end{align}
for $\ell\in\cu{0,1,2}$.  Note that $I_r=I_0$ and $I_\sigma=I_2$.  The radial trajectory satisfies the squared differential equation
\begin{align}
	\label{eq:RadialODE}
	\pa{\frac{dr_o}{d\tau}}^2=\mathcal{R}(r_o(\tau)),
\end{align}
while the radial integrals \eqref{eq:EllipticIntegralsBasis} satisfy
\begin{align}
	\label{eq:IkODE}
	\frac{dI_\ell}{d\tau}=\br{r_o(\tau)}^\ell,\qquad
	\frac{dI_\pm}{d\tau}=\frac{1}{r_o(\tau)-r_\pm}.
\end{align}

Below we will go through each case (1)-(4) [Sec.~\ref{sec:RadialIntegrals}], following the basic approach established in our treatment of the angular integrals in Sec.~\ref{sec:AngularPotential} above.  Each case requires a different substitution of variables to obtain real and smooth antiderivatives for the integrals.\footnote{Near critical values of the roots, the integrals may be approximated using the method of matched asymptotic expansions \cite{Gralla2017,KerrLensing}.}  Many of the relevant substitutions are summarized in \S3.145--3.151 of Ref.~\cite{Gradshteyn2007}, and full details are presented in \S250--267 of Ref.~\cite{Byrd1954}.  The antiderivatives can be used to calculate the path integrals as functions of $r_s$, $r_o$, $\nu_r$, and the number of turning points $w$ encountered along the trajectory via
\begin{align}
	I_i=(-1)^w\mathcal{I}_i^o-\mathcal{I}_i^s,
\end{align}
where $\mathcal{I}_i^j$ indicates the antiderivative $\mathcal{I}_i$ evaluated at the source ($j=s$) or observer ($j=o$).  Here, we assumed that the trajectory encounters either no turning points ($w=0$) or a single turning point ($w=1$), which is appropriate for our restriction to motion in the black hole exterior.

\subsection*{Preliminaries}

For the analysis that follows, it is most convenient to think of the radial potential as a quartic in terms of its roots,
\begin{align}
	\mathcal{R}(r)=\pa{r-r_1}\pa{r-r_2}\pa{r-r_3}\pa{r-r_4}.
\end{align}
Throughout this section, we use the notation
\begin{align}
	r_{ij}=r_i-r_j,
\end{align}
with $i,j\in\cu{1,2,3,4,+,-}$ to represent the four roots of the radial potential, as well as the inner and outer horizons.  In addition, if $r_3=\bar{r}_4$ are complex conjugates, then it is useful to set
\begin{align}
	\label{eq:a1b1}
	r_3=b_1-ia_1,\qquad
	r_4=b_1+ia_1,\qquad
	a_1=\sqrt{-\frac{r_{43}^2}{4}}
	>0,\qquad
	b_1=\frac{r_3+r_4}{2}
	=z
	>0,
\end{align}
where the last equality follows from the explicit expressions \eqref{eq:RadialRoots} for the radial roots, and the following inequality from Eq.~\eqref{eq:z}.  Moreover, if $r_1=\bar{r}_2$ are also complex conjugates, then we likewise set
\begin{align}
	\label{eq:a2b2}
	r_1=b_2-ia_2,\qquad
	r_2=b_2+ia_2,\qquad
	a_2=\sqrt{-\frac{r_{21}^2}{4}}
	>0,\qquad
	b_2=\frac{r_1+r_2}{2}
	=-z<0.
\end{align}
Note that the condition $r_1+r_2+r_3+r_4=0$ is automatically enforced, as expected from Eq.~\eqref{eq:Vieta}, and also that
\begin{align}
	\label{eq:ComplexRootOrdering}
	b_1>0>b_2=-b_1,\qquad
	a_1>a_2>0.
\end{align}
Lastly, we define once and for all the elliptic parameter
\begin{align}
	\label{eq:EllipticParameter}
	k=\frac{r_{32}r_{41}}{r_{31}r_{42}}.
\end{align}
Note that if all the roots are real, then their ordering ensures that $r_{32}r_{41}-r_{31}r_{42}=-r_{43}r_{21}<0$, and hence that
\begin{align}
	k\in(0,1).
\end{align}
On the other hand, if two of the roots are complex, then $k\in\mathbb{C}$ is a pure complex phase, as can be seen by plugging in Eqs.~\eqref{eq:a1b1} into the definition \eqref{eq:EllipticParameter}.  When all the roots are complex, then $k>1$, as can be seen by plugging in Eqs.~\eqref{eq:a1b1} and \eqref{eq:a2b2} into the definition \eqref{eq:EllipticParameter}.  In these cases, another parameter is needed instead.

\subsection{Case (1)}

In case (1), all four roots are real and the range of radial motion is $r_1<r_2\le r\le r_3<r_4$.  (In the maximally extended spacetime, the photon alternates between successive universes.)  An appropriate substitution for the evaluation of the integrals \eqref{eq:EllipticIntegralsBasis} is then (see Eq.~(4) in \S3.147 of Ref.~\cite{Gradshteyn2007} and \S254 of Ref.~\cite{Byrd1954})
\begin{align}
	\label{eq:x1}
	x_1(r)=\sqrt{\frac{r-r_2}{r-r_1}\frac{r_{31}}{r_{32}}}
	\in(0,1].
\end{align}
The parameter \eqref{eq:EllipticParameter} is less than unity, $k\in(0,1)$, and the antiderivatives
\begin{align}
	\label{eq:IrCase1}
	\mathcal{I}_0&=F^{(1)}(r),\\
	\mathcal{I}_1&=r_1F^{(1)}(r)+r_{21}\Pi_1^{(1)}(r),\\
	\mathcal{I}_2&=\frac{\sqrt{\mathcal{R}(r)}}{r-r_1}-\frac{r_1r_4+r_2r_3}{2}F^{(1)}(r)-E^{(1)}(r),\\
	\mathcal{I}_\pm&=-\Pi_\pm^{(1)}(r)-\frac{F^{(1)}(r)}{r_{\pm1}},
\end{align}
are real and smooth, with\footnote{$F^{(1)}\ge0$, $E^{(1)}\ge0$, and $\Pi_1^{(1)}\ge0$ because $F(\varphi|k)\ge0$, $E(\varphi|k)\ge0$ and $\Pi(n;\varphi|k)\ge0$ whenever $\varphi\in\br{0,\frac{\pi}{2}}$, $k\in\br{0,1}$, and $n\in\br{0,1}$.  On the other hand, $\Pi_\pm^{(1)}$ can be negative.}
\begin{align}
	\label{eq:F1}
	F^{(1)}(r)&=\frac{2}{\sqrt{r_{31}r_{42}}}F\pa{\arcsin{x_1(r)}\Big|k}
	\ge0,\\
	E^{(1)}(r)&=\sqrt{r_{31}r_{42}}E\pa{\arcsin{x_1(r)}\Big|k}
	\ge0,\\
	\Pi_1^{(1)}(r)&=\frac{2}{\sqrt{r_{31}r_{42}}}\Pi\pa{\frac{r_{32}}{r_{31}};\arcsin{x_1(r)}\bigg|k}
	\ge0,\\
	\Pi_\pm^{(1)}(r)&=\frac{2}{\sqrt{r_{31}r_{42}}}\frac{r_{21}}{r_{\pm1}r_{\pm2}}\Pi\pa{\frac{r_{\pm1}r_{32}}{r_{\pm2}r_{31}};\arcsin{x_1(r)}\bigg|k},
\end{align}
In general, the expression for $\mathcal{I}_2$ contains an additional contribution from the antiderivative
\begin{align}
	\frac{r_1+r_2+r_3+r_4}{2}\pa{r_1F^{(1)}(r)+r_{21}\Pi_1^{(1)}(r)}=0,
\end{align}
which in this case vanishes by Eq.~\eqref{eq:Vieta}.

\subsubsection{Inversion for \texorpdfstring{$r_o(\tau)$}{radial motion}}

Before a turning point is reached, the path integral $I_r=\tau$ beginning from $r=r_s$ is given by
\begin{align}
	\label{eq:tauIr}
	\tau=I_r
	=\nu_r\pa{\mathcal{I}_r^o-\mathcal{I}_r^s},\qquad
	\nu_r=\sign\pa{p_s^r}.
\end{align}
As in the angular analysis of Sec.~\ref{sec:AngularPotential}, we may use the Jacobi elliptic sine function to invert this equation.  Recalling that $\mathcal{I}_r=\mathcal{I}_0=F^{(1)}(r)$,  Eqs.~\eqref{eq:JacobiSN}, \eqref{eq:F1}, and \eqref{eq:tauIr} can be combined to give
\begin{align}
	\label{eq:X1}
	x_1(r_o)=\nu_r\sn\pa{X_1(\tau)\big|k},\qquad
	X_1(\tau)=\frac{\sqrt{r_{31}r_{42}}}{2}\pa{\tau+\nu_r\mathcal{I}_r^s}.
\end{align}
Solving for $r_o$ using Eq.~\eqref{eq:x1}, we then find
\begin{align}
	\label{eq:Inversion1}
	r_o^{(1)}(\tau)=\frac{r_2r_{31}-r_1r_{32}\sn^2\pa{X_1(\tau)\big|k}}{r_{31}-r_{32}\sn^2\pa{X_1(\tau)\big|k}},
\end{align}
where we write $r_o^{(1)}$ to emphasize that this formula for $r_o(\tau)$ was derived in case (1), even though it will extend to the other cases with little modification.  Although Eq.~\eqref{eq:Inversion1} was derived under the assumption that a turning point has not yet been reached, it in fact continues properly through turning points to provide the full parameterized trajectory $r_o(\tau)$, as follows.  Noting that $\sn^2(\varphi|k)$ oscillates smoothly between $0$ and $+1$ with period $2K(k)$, we see that Eq.~\eqref{eq:Inversion1} defines a trajectory $r_o(\tau)$ that oscillates between $r_2$ and $r_3$ with period $2K(k)$.  Thus, it has the correct quantitative behavior at turning points, and we need only check that it satisfies the squared differential equation \eqref{eq:RadialODE}, which is easily verified using the elliptic identities \eqref{eq:JacobiSNCN} and \eqref{eq:JacobiSNDN}.  This completes the proof that Eq.~\eqref{eq:Inversion1} is the unique solution for $r_o(\tau)$ with initial conditions $r_o(0)=r_s$ and $\sign\br{r_o'(0)}=\nu_r$, which is manifestly real in case (1).

\subsubsection{Path integrals as a function of Mino time}

We have $I_r=I_0=\tau$ by definition, and the other path integrals may be expressed in terms of Mino time $\tau$ using a slight extension of the method used in Sec.~\ref{sec:AngularPotential} for the angular case.  Before the first turning point is reached, we have
\begin{align}
	\label{eq:DefiniteI}
	I_i=\nu_r\pa{\mathcal{I}_i^o-\mathcal{I}_i^s},
\end{align}
where the antiderivatives $\mathcal{I}_i$ depend on $r$ primarily through the combination $\arcsin{x_1(r)}$.  Applying the formula \eqref{eq:JacobiAmplitude} to Eq.~\eqref{eq:X1} extends $\arcsin{x_1(r)}$ to the (monotonically increasing in $\tau$) amplitude
\begin{align}
	\label{eq:x1Amplitude}
	\arcsin{x_1(r)}=\nu_r\am\pa{X_1(\tau)\big|k},
\end{align}
where $X_1(\tau)$ is as defined in Eq.~\eqref{eq:X1}, and the restriction $|\varphi|\le K(k)$ is satisfied on account of our assumption that a turning point has not yet been reached.  Plugging the extension \eqref{eq:x1Amplitude} into Eq.~\eqref{eq:DefiniteI} as needed, we then find
\begin{subequations}
\label{eq:RadialCase2}
\begin{align}
	I_1&=r_1\tau+r_{21}\Pi_{1,\tau}^{(1)},\\
	I_2&=H_\tau^{(1)}-\frac{r_1r_4+r_2r_3}{2}\tau-E_\tau^{(1)},\\
	I_\pm&=-\frac{\tau}{r_{\pm1}}-\Pi_{\pm,\tau}^{(1)},
\end{align}
\end{subequations}
with
\begin{align}
	H_\tau^{(1)}&=\frac{r_o'(\tau)}{r_o(\tau)-r_1}-\nu_r\frac{\sqrt{\mathcal{R}(r_s)}}{r_s-r_1},\\
	E_\tau^{(1)}&=\sqrt{r_{31}r_{42}}\br{E\pa{\am\pa{X_1(\tau)\big|k}\Big|k}-\nu_rE\pa{\arcsin{x_1(r_s)}\Big|k}},\\
	\Pi_{1,\tau}^{(1)}&=\frac{2}{\sqrt{r_{31}r_{42}}}\br{\Pi\pa{\frac{r_{32}}{r_{31}};\am\pa{X_1(\tau)\big|k}\bigg|k}-\nu_r\Pi\pa{\frac{r_{32}}{r_{31}};\arcsin{x_1(r_s)}\bigg|k}},\\
	\Pi_{\pm,\tau}^{(1)}&=\frac{2}{\sqrt{r_{31}r_{42}}}\frac{r_{21}}{r_{\pm1}r_{\pm2}}\br{\Pi\pa{\frac{r_{\pm1}r_{32}}{r_{\pm2}r_{31}};\am\pa{X_1(\tau)\big|k}\bigg|k}-\nu_r\Pi\pa{\frac{r_{\pm1}r_{32}}{r_{\pm2}r_{31}};\arcsin{x_1(r_s)}\bigg|k}}.
\end{align}
These functions of Mino time vanish at $\tau=0$ by construction.

Although these expressions were derived under the assumption that a turning point does not occur, they in fact extend properly through turning points to give the complete path integrals $I_i$, as follows.  Since $\am(\varphi|k)$, $F(\varphi|k)$, $E(\varphi|k)$, and $\Pi(n;\varphi|k)$ are real and smooth functions of $\varphi$ provided $\max(k,n)<1$ (which is the case here), it follows that the candidate formulas for the $I_i$ are real and smooth.  Thus, we need only check the differential equations \eqref{eq:IkODE}, which is straightforward using Eq.~\eqref{eq:x1Amplitude} together with the quasiperiodicity properties \eqref{eq:QuasiPeriodicity}.  This completes the proof that these formulas give the full path integrals \eqref{eq:EllipticIntegralsBasis}.

\subsection{Case (2)}

In case (2), all four roots are real and the range of radial motion is $r_1<r_2<r_3<r_4\le r$.  An appropriate substitution for the evaluation of the integrals \eqref{eq:EllipticIntegralsBasis} is then (see Eq.~(8) in \S3.147 of Ref.~\cite{Gradshteyn2007} and \S258 of Ref.~\cite{Byrd1954})
\begin{align}
	\label{eq:x2}
	x_2(r)=\sqrt{\frac{r-r_4}{r-r_3}\frac{r_{31}}{r_{41}}}
	\in\br{0,\sqrt{\frac{r_{31}}{r_{41}}}}
	\subset[0,1).
\end{align}
The parameter \eqref{eq:EllipticParameter} is less than unity, $k\in(0,1)$, and the antiderivatives
\begin{align}
	\label{eq:IrCase2}
	\mathcal{I}_0&=F^{(2)}(r),\\
	\mathcal{I}_1&=r_3F^{(2)}(r)+r_{43}\Pi_1^{(2)}(r),\\
	\mathcal{I}_2&=\frac{\sqrt{\mathcal{R}(r)}}{r-r_3}-\frac{r_1r_4+r_2r_3}{2}F^{(2)}(r)-E^{(2)}(r),\\
	\mathcal{I}_\pm&=-\Pi_\pm^{(2)}(r)-\frac{F^{(2)}(r)}{r_{\pm3}},
\end{align}
are real and smooth, with\footnote{$F^{(2)}\ge0$, $E^{(2)}\ge0$, and $\Pi_1^{(2)}\ge0$ because $F(\varphi|k)\ge0$, $E(\varphi|k)\ge0$ and $\Pi(n;\varphi|k)\ge0$ whenever $\varphi\in\br{0,\arcsin{\frac{1}{\sqrt{n}}}}$, $k\in\br{0,1}$, and $n\ge1$.  On the other hand, $\Pi_\pm^{(2)}$ can be negative.}
\begin{align}
	\label{eq:F2}
	F^{(2)}(r)&=\frac{2}{\sqrt{r_{31}r_{42}}}F\pa{\arcsin{x_2(r)}\Big|k}
	\ge0,\\
	E^{(2)}(r)&=\sqrt{r_{31}r_{42}}E\pa{\arcsin{x_2(r)}\Big|k}
	\ge0,\\
	\Pi_1^{(2)}(r)&=\frac{2}{\sqrt{r_{31}r_{42}}}\Pi\pa{\frac{r_{41}}{r_{31}};\arcsin{x_2(r)}\bigg|k}
	\ge0,\\
	\Pi_\pm^{(2)}(r)&=\frac{2}{\sqrt{r_{31}r_{42}}}\frac{r_{43}}{r_{\pm3}r_{\pm4}}\Pi\pa{\frac{r_{\pm3}r_{41}}{r_{\pm4}r_{31}};\arcsin{x_2(r)}\bigg|k},
\end{align}
In general, the expression for $\mathcal{I}_2$ contains an additional contribution from the antiderivative
\begin{align}
	\frac{r_1+r_2+r_3+r_4}{2}\pa{r_3F^{(2)}(r)+r_{43}\Pi_1^{(2)}(r)}=0,
\end{align}
which in this case vanishes by Eq.~\eqref{eq:Vieta}.

\subsubsection{Inversion for \texorpdfstring{$r_o(\tau)$}{radial motion}}

Before a turning point is reached, the path integral $I_r=\tau$ beginning from $r=r_s$ is still given by Eq.~\eqref{eq:tauIr}.  As usual, we may use the Jacobi elliptic sine function to invert this equation.  Recalling that $\mathcal{I}_r=\mathcal{I}_0=F^{(2)}(r)$,  Eqs.~\eqref{eq:JacobiSN}, \eqref{eq:tauIr}, and \eqref{eq:F2} can be combined to give
\begin{align}
	\label{eq:X2}
	x_2(r_o)=\nu_r\sn\pa{X_2(\tau)\big|k},\qquad
	X_2(\tau)=\frac{\sqrt{r_{31}r_{42}}}{2}\pa{\tau+\nu_r\mathcal{I}_r^s}.
\end{align}
Solving for $r_o$ using Eq.~\eqref{eq:x2}, we then find
\begin{align}
	\label{eq:Inversion2}
	r_o^{(2)}(\tau)=\frac{r_4r_{31}-r_3r_{41}\sn^2\pa{X_2(\tau)\big|k}}{r_{31}-r_{41}\sn^2\pa{X_2(\tau)\big|k}}.
\end{align}
where we write $r_o^{(2)}$ to emphasize that this formula for $r_o(\tau)$ was derived in case (2), even though it will extend to the other cases with little modification.  Although Eq.~\eqref{eq:Inversion2} was derived under the assumption that a turning point has not yet been reached, it in fact continues properly through the turning point $r_4$ (if encountered) to provide the full parameterized trajectory $r_o(\tau)$.  (If $\nu_r<0$ there is no turning point in the future of the initial data.)  Noting that $\sn^2(\varphi|k)$ oscillates smoothly between $0$ and $+1$ with period $2K(k)$, we see that Eq.~\eqref{eq:Inversion2} defines a trajectory $r_o(\tau)$ that correctly bounces when $r=r_4$, where $\sn^2(\varphi|k)=0$.  Thus, it has the correct quantitative behavior at the turning point, and we need only check that it satisfies the squared differential equation \eqref{eq:RadialODE}, which is easily verified using the elliptic identities \eqref{eq:JacobiSNCN} and \eqref{eq:JacobiSNDN}.  This completes the proof that Eq.~\eqref{eq:Inversion2} is the unique solution for $r_o(\tau)$ with initial conditions $r_o(0)=r_s$ and $\sign\br{r_o'(0)}=\nu_r$, which is manifestly real in case (2).

\subsubsection{Path integrals as a function of Mino time}

We have $I_r=I_0=\tau$ by definition, and the other path integrals may be expressed in terms of Mino time $\tau$ using the same method as usual.  Before the first turning point is reached, we once again have Eq.~\eqref{eq:DefiniteI}, where the antiderivatives $\mathcal{I}_i$ depend on $r$ primarily through the combination $\arcsin{x_2(r)}$.  Applying the formula \eqref{eq:JacobiAmplitude} to Eq.~\eqref{eq:X2} extends $\arcsin{x_2(r)}$ to the (monotonically increasing in $\tau$) amplitude
\begin{align}
	\label{eq:x2Amplitude}
	\arcsin{x_2(r)}=\nu_r\am\pa{X_2(\tau)\big|k},
\end{align}
where $X_2(\tau)$ is as defined in Eq.~\eqref{eq:X2}, and the restriction $|\varphi|\le K(k)$ is satisfied on account of our assumption that a turning point has not yet been reached.  Plugging the extension \eqref{eq:x2Amplitude} into Eq.~\eqref{eq:DefiniteI} as needed, we then find
\begin{align}
	I_1&=r_3\tau+r_{43}\Pi_{1,\tau}^{(2)},\\
	I_2&=H_\tau^{(2)}-\frac{r_1r_4+r_2r_3}{2}\tau-E_\tau^{(2)},\\
	I_\pm&=-\frac{\tau}{r_{\pm3}}-\Pi_{\pm,\tau}^{(2)},
\end{align}
with
\begin{align}
	H_\tau^{(2)}&=\frac{r_o'(\tau)}{r_o(\tau)-r_3}-\nu_r\frac{\sqrt{\mathcal{R}(r_s)}}{r_s-r_3},\\
	E_\tau^{(2)}&=\sqrt{r_{31}r_{42}}\br{E\pa{\am\pa{X_2(\tau)\big|k}\Big|k}-\nu_rE\pa{\arcsin{x_2(r_s)}\Big|k}},\\
	\Pi_{1,\tau}^{(2)}&=\frac{2}{\sqrt{r_{31}r_{42}}}\br{\Pi\pa{\frac{r_{41}}{r_{31}};\am\pa{X_2(\tau)\big|k}\bigg|k}-\nu_r\Pi\pa{\frac{r_{41}}{r_{31}};\arcsin{x_2(r_s)}\bigg|k}},\\
	\Pi_{\pm,\tau}^{(1)}&=\frac{2}{\sqrt{r_{31}r_{42}}}\frac{r_{43}}{r_{\pm3}r_{\pm4}}\br{\Pi\pa{\frac{r_{\pm3}r_{41}}{r_{\pm4}r_{31}};\am\pa{X_2(\tau)\big|k}\bigg|k}-\nu_r\Pi\pa{\frac{r_{\pm3}r_{41}}{r_{\pm4}r_{31}};\arcsin{x_2(r_s)}\bigg|k}}.
\end{align}
These functions of Mino time vanish at $\tau=0$ by construction.

Although these expressions were derived under the assumption that a turning point does not occur, they in fact extend properly through the turning point $r_4$ (if encountered) to give the complete path integrals $I_i$, as follows.  Since $\am(\varphi|k)$, $F(\varphi|k)$, $E(\varphi|k)$, and $\Pi(n;\varphi|k)$ are real and smooth functions of $\varphi$ provided $\max(k,n)<1$ (which is the case here), it follows that the candidate formulas for the $I_i$ are real and smooth.  Thus, we need only check the differential equations \eqref{eq:IkODE}, which is straightforward using Eq.~\eqref{eq:x2Amplitude} together with the quasiperiodicity properties \eqref{eq:QuasiPeriodicity}.  This completes the proof that these formulas give the full path integrals \eqref{eq:EllipticIntegralsBasis}.

\subsection{Case (3)}

In case (3), only two roots are real and the range of radial motion is $r_1<r_2<r_-<r_+\le r_i$ with $r_3=\bar{r}_4$.  An appropriate substitution for the evaluation of the integrals \eqref{eq:EllipticIntegralsBasis} is then (see Eq.~(1) in \S3.145 of Ref.~\cite{Gradshteyn2007} and \S260 of Ref.~\cite{Byrd1954})\footnote{Here, the two references superficially disagree, but they are in fact related by the identity $\arccos{x}=2\arctan{\frac{\sqrt{1-x^2}}{1+x}}$, valid for $x\in(-1,1]$.}
\begin{align}
	\label{eq:x3}
	x_3(r)=\frac{A\pa{r-r_1}-B\pa{r-r_2}}{A\pa{r-r_1}+B\pa{r-r_2}},
\end{align}
where [recall Eq.~\eqref{eq:a1b1}]
\begin{align}
	A^2=a_1^2+\pa{b_1-r_2}^2
	>0,\qquad
	B^2=a_1^2+\pa{b_1-r_1}^2
	>0,
\end{align}
and we must choose the same sign for the square root in $A$ and $B$.  Picking the positive branch results in
\begin{align}
	A=\sqrt{r_{32}r_{42}}
	>0,\qquad
	B=\sqrt{r_{31}r_{41}}
	>0.
\end{align}
Moreover, since $r_3+r_4=2z>0$ [Eq.~\eqref{eq:a1b1}] and $r_1+r_2=-2z<0$ [Eq.~\eqref{eq:Vieta}], it follows that $r_3+r_4>r_1+r_2$, and hence that $B^2-A^2=r_{21}\pa{r_3+r_4-r_1-r_2}>0$.  Therefore, using the fact that $r_1<r_2<r$,
\begin{align}
	\label{eq:x_3}
	B>A>0,\qquad
	\alpha_0=\frac{B+A}{B-A}>1,\qquad
	x_3(r)=\frac{1-\frac{B\pa{r-r_2}}{A\pa{r-r_1}}}{1+\frac{B\pa{r-r_2}}{A\pa{r-r_1}}}
	\in\pa{-\frac{1}{\alpha_0},1}
	\subset(-1,1).
\end{align}
While in this case, the parameter \eqref{eq:EllipticParameter} is a pure phase $k\in\mathbb{C}$, we can replace it by a new parameter that is real, positive, and less than unity:
\begin{align}
	\label{eq:EllipticParameter3}
	k_3=\frac{\pa{A+B}^2-r_{21}^2}{4AB}
	=\frac{1}{2}\br{1+\frac{a_1^2+\pa{b_1-r_1}\pa{b_1-r_2}}{\sqrt{\br{a_1^2+\pa{b_1-r_1}\pa{b_1-r_2}}^2+a_1^2r_{21}^2}}}
	\in(0,1).
\end{align}
We now invoke the results in \S341 of Ref.~\cite{Byrd1954} (though note that we use different conventions for the elliptic integrals).  After correcting an error in the auxiliary formula \S361.54 (their $f_1$ is missing a factor of $\frac{1}{2}$), we find that whenever
\begin{align}
	\label{eq:ConditionsForR}
	j\in\pa{0,1},\qquad
	\alpha^2>1,\qquad
	\varphi\in\left[0,\pi-\arccos{\frac{1}{\alpha}}\right),
\end{align}
so that $\alpha^2/\pa{\alpha^2-1}>j$ is automatically satisfied, then
\begin{align}
	R_1\pa{\alpha;\varphi\big|j}&=\int_0^{F(\varphi|j)}\frac{\ed u}{1+\alpha\cn(u|j)}
	=\int_0^\varphi\frac{\ed t}{\pa{1+\alpha\cos{t}}\sqrt{1-j\sin^2{t}}}\\
	&=\frac{1}{1-\alpha^2}\br{\Pi\pa{\left.\frac{\alpha^2}{\alpha^2-1};\varphi\right|j}-\alpha f_1},\\
	R_2\pa{\alpha;\varphi|j}&=\int_0^{F(\varphi|j)}\frac{\ed u}{\br{1+\alpha\cn(u|j)}^2}
	=\int_0^\varphi\frac{\ed t}{\pa{1+\alpha\cos{t}}^2\sqrt{1-j\sin^2{t}}}\\
	&=\frac{1}{\alpha^2-1}\br{F\pa{\varphi\big|j}-\frac{\alpha^2}{j+\pa{1-j}\alpha^2}\pa{E\pa{\varphi\big|j}-\frac{\alpha\sin{\varphi}\sqrt{1-j\sin^2{\varphi}}}{1+\alpha\cos{\varphi}}}}\nonumber\\
	&\qquad
	+\frac{1}{j+\pa{1-j}\alpha^2}\pa{2j-\frac{\alpha^2}{\alpha^2-1}}R_1\pa{\alpha;\varphi\big|j},\\
	f_1&=\frac{p_1}{2}\log\ab{\frac{p_1\sqrt{1-j\sin^2{\varphi}}+\sin{\varphi}}{p_1\sqrt{1-j\sin^2{\varphi}}-\sin{\varphi}}}
	\ge0,\qquad
	p_1=\sqrt{\frac{\alpha^2-1}{j+\pa{1-j}\alpha^2}}
	>0.
\end{align}

Finally, we also introduce the parameters
\begin{align}
	\alpha_\pm=\frac{Br_{\pm2}+Ar_{\pm1}}{Br_{\pm2}-Ar_{\pm1}}
	=-\frac{1}{x_3(r_\pm)}.
\end{align}
Since the conditions \eqref{eq:ConditionsForR} are satisfied by $j=k_3$, $\alpha=\alpha_{0,\pm}$, and $\varphi=\arccos{x_3(r)}$, the antiderivatives
\begin{align}
	\label{eq:IrCase3}
	\mathcal{I}_0&=F^{(3)}(r),\\
	\mathcal{I}_1&=\pa{\frac{Br_2+Ar_1}{B+A}}F^{(3)}(r)+\Pi_1^{(3)}(r),\\
	\mathcal{I}_2&=\pa{\frac{Br_2+Ar_1}{B+A}}^2F^{(3)}(r)+2\pa{\frac{Br_2+Ar_1}{B+A}}\Pi_1^{(3)}(r)+\sqrt{AB}\Pi_2^{(3)}(r),\\
	\mathcal{I}_\pm&=-\frac{1}{Br_{\pm2}+Ar_{\pm1}}\br{\pa{B+A}F^{(3)}(r)+\frac{2r_{21}\sqrt{AB}}{Br_{\pm2}-Ar_{\pm1}}R_1\pa{\alpha_\pm;\arccos{x_3(r)}\Big|k_3}},
\end{align}
are real and smooth, with\footnote{$F^{(3)}>0$ and $\Pi_k^{(3)}>0$ because $F(\varphi|j)\ge0$ and $R_k(\alpha;\varphi|j)\ge0$ whenever $\varphi\in\br{0,\pi-\arccos{\frac{1}{\alpha}}}$, $j\in\pa{0,1}$, and $\alpha>1$ (in the case of $R_k$, this is manifest from the integrand).  Note however that if $\alpha<-1$, then $R_1\le0$ changes sign while $R_2\ge0$ does not.}
\begin{align}
	\label{eq:F3}
	F^{(3)}(r)&=\frac{1}{\sqrt{AB}}F\pa{\arccos{x_3(r)}\Big|k_3}
	>0,\\
	\Pi_\ell^{(3)}(r)&=\pa{\frac{2r_{21}\sqrt{AB}}{B^2-A^2}}^\ell R_\ell\pa{\alpha_0;\arccos{x_3(r)}\Big|k_3}
	>0.
\end{align}

\subsubsection{Inversion for \texorpdfstring{$r_o(\tau)$}{radial motion}}

Before a turning point is reached, the path integral $I_r=\tau$ beginning from $r=r_s$ is still given by Eq.~\eqref{eq:tauIr}, which in this case can be inverted using the Jacobi elliptic cosine function $\cn(\varphi|k)$.  This function satisfies
\begin{align}
	\label{eq:JacobiCN}
	\cn\pa{F(\arccos{\varphi}|k)}=\varphi,
\end{align}
and is even in its first argument, $\cn(-\varphi|k)=\cn(\varphi|k)$.  Recalling that $\mathcal{I}_r=\mathcal{I}_0=F^{(3)}(r)$,  Eqs.~\eqref{eq:tauIr}, \eqref{eq:F3}, and \eqref{eq:JacobiCN} can be combined to give
\begin{align}
	\label{eq:X3}
	x_3(r_o)=\cn\pa{X_3(\tau)\big|k_3},\qquad
	X_3(\tau)=\sqrt{AB}\pa{\tau+\nu_r\mathcal{I}_r^s}.
\end{align}
Solving for $r_o$ using Eq.~\eqref{eq:x3}, we then find
\begin{align}
	\label{eq:Inversion3}
	r_o^{(3)}(\tau)=\frac{\pa{Br_2-Ar_1}+\pa{Br_2+Ar_1}\cn\pa{X_3(\tau)\big|k_3}}{\pa{B-A}+\pa{B+A}\cn\pa{X_3(\tau)\big|k_3}},
\end{align}
where we write $r_o^{(3)}$ to emphasize that this formula for $r_o(\tau)$ was derived in case (3), even though it will extend to the other cases with little modification.  This trajectory never encounters a turning point outside the horizon, and hence Eq.~\eqref{eq:Inversion3} is the unique solution for $r_o(\tau)$ with initial conditions $r_o(0)=r_s$ and $\sign\br{r_o'(0)}=\nu_r$, which is manifestly real in case (3).

\subsubsection{Path integrals as a function of Mino time}

We have $I_r=I_0=\tau$ by definition, and the other path integrals may be expressed in terms of Mino time $\tau$ using the same method as usual.  Since there are no turning points, we once again have Eq.~\eqref{eq:DefiniteI}, where the antiderivatives $\mathcal{I}_i$ depend on $r$ primarily through the combination $\arccos{x_3(r)}$.  We now invoke the inversion formula
\begin{align}
	\am(\varphi|k)=\arccos\pa{\cn(\varphi|k)},\qquad
	0\le\varphi\le2K(k).
\end{align}
Applying it to Eq.~\eqref{eq:X3} extends $\arccos{x_3(r)}$ to the (monotonically increasing in $\tau$) amplitude
\begin{align}
	\label{eq:x3Amplitude}
	\arccos{x_3(r)}=\am\pa{X_3(\tau)\big|k},
\end{align}
where $X_3(\tau)$ is as defined in Eq.~\eqref{eq:X3}, and the restriction $0\le\varphi\le2K(k)$ is satisfied on the whole range of motion.  Plugging the extension \eqref{eq:x3Amplitude} into Eq.~\eqref{eq:DefiniteI} as needed, we then find
\begin{align}
	I_1&=\pa{\frac{Br_2+Ar_1}{B+A}}\tau+\Pi_{1,\tau}^{(3)},\\
	I_2&=\pa{\frac{Br_2+Ar_1}{B+A}}^2\tau+2\pa{\frac{Br_2+Ar_1}{B+A}}\Pi_{1,\tau}^{(3)}+\sqrt{AB}\Pi_{2,\tau}^{(3)},\\
	I_\pm&=-\frac{\pa{B+A}\tau+\Pi_{\pm,\tau}^{(3)}}{Br_{\pm2}+Ar_{\pm1}},
\end{align}
with
\begin{align}
	\Pi_{\ell,\tau}^{(3)}&=\pa{\frac{2r_{21}\sqrt{AB}}{B^2-A^2}}^\ell\br{R_\ell\pa{\alpha_0;\am\pa{X_3(\tau)\big|k_3}\Big|k_3}-\nu_rR_\ell\pa{\alpha_0;\arccos{x_3(r_s)}\Big|k_3}},\\
	\Pi_{\pm,\tau}^{(3)}&=\pa{\frac{2r_{21}\sqrt{AB}}{Br_{\pm2}-Ar_{\pm1}}}\br{R_1\pa{\alpha_\pm;\am\pa{X_3(\tau)\big|k_3}\Big|k_3}-\nu_rR_1\pa{\alpha_\pm;\arccos{x_3(r_s)}\Big|k_3}}.
\end{align}
These functions of Mino time vanish at $\tau=0$ by construction.  Since a turning point never occurs, these formulas give the full path integrals \eqref{eq:EllipticIntegralsBasis}.

\subsection{Case (4)}

In case (4), there are two pairs $r_1=\bar{r}_2$ and $r_3=\bar{r}_4$ of complex conjugate roots and the range of radial motion is unbounded.  An appropriate substitution for the evaluation of the integrals \eqref{eq:EllipticIntegralsBasis} is then (see Eq.~(4) in \S3.145 of Ref.~\cite{Gradshteyn2007} and \S267 of Ref.~\cite{Byrd1954})\footnote{Here, the two references superficially disagree, but they are in fact related by the arctangent addition formula. However, the use of this formula in Ref.~\cite{Byrd1954} introduces a discontinuity in the antiderivative, so we instead use its always continuous form given in Ref.~\cite{Gradshteyn2007}.}
\begin{align}
	\label{eq:x4}
	x_4(r)=\frac{r-b_2}{a_2}
	=\frac{r+b_1}{a_2}
	>0,
\end{align}
where we used the fact that $b_2=-b_1=-z<0$ [Eq.~\eqref{eq:a2b2}], so that $r>0>-b_1$ outside the horizon; since $a_2>0$ [Eq.~\eqref{eq:ComplexRootOrdering}], this guarantees that $x_4(r)$ is positive.  In principle, we could have equivalently defined $x_4(r)=\pa{r-b_1}/a_1$, but then we would also have had to consider negative values of $x_4(r)$. The present choice will prove more convenient.

Following Refs.~\cite{Byrd1954,Gradshteyn2007}, we must also introduce the quantities [recall Eqs.~\eqref{eq:a1b1} and \eqref{eq:a2b2}]
\begin{align}
	C^2=\pa{a_1-a_2}^2+\pa{b_1-b_2}^2
	>0,\qquad
	D^2=\pa{a_1+a_2}^2+\pa{b_1-b_2}^2
	>0,
\end{align}
which are well-defined up to a choice of sign in the square root.  Picking the positive branch results in
\begin{align}
	C=\sqrt{r_{31}r_{42}}
	>0,\qquad
	D=\sqrt{r_{32}r_{41}}
	>0,
\end{align}
from which it then follows that
\begin{align}
	k=\frac{D^2}{C^2}
	=1+\frac{4a_1a_2}{\pa{a_1-a_2}^2+\pa{b_1-b_2}^2}
	>1.
\end{align}
While the parameter $k>1$, we can replace it by a new elliptic parameter that is real, positive, and less than unity:
\begin{align}
	k_4=\frac{4CD}{\pa{C+D}^2}
	=\frac{4\sqrt{k}}{\pa{1+\sqrt{k}}^2}
	\in(0,1).
\end{align}
The reduction of the elliptic integrals \eqref{eq:EllipticIntegralsBasis} to Legendre normal form presented in Refs.~\cite{Byrd1954,Gradshteyn2007} further requires\footnote{This definition agrees with Ref.~\cite{Gradshteyn2007} ($g_0=\tan{\alpha}$) and Ref.~\cite{Byrd1954} ($g_0^2=g_1^2$), except that $a_2$ replaces $a_1$ because our $x_4(r)$ is $\pa{r-b_2}/a_2$ instead of $\pa{r-b_1}/a_1$.  Unfortunately, $g_0=\pm\sqrt{g_0^2}$ exhibits a sign ambiguity in the choice of branch for the square root.  This has led some authors such as Dexter \& Agol \cite{Dexter2009} to prefer the use of Carlson's symmetric integrals, which notably do not suffer from this sign ambiguity \cite{Carlson1992}.  However, in the present case, picking the positive branch in the formulas of Ref.~\cite{Byrd1954} always yields the correct answer.}
\begin{align}
	g_0=\sqrt{\frac{4a_2^2-\pa{C-D}^2}{\pa{C+D}^2-4a_2^2}}
	\in(0,1),
\end{align}
with the last inequality rendered manifest by the relations [recall from Eq.~\eqref{eq:ComplexRootOrdering} that $a_1>a_2>0$]
\begin{align}
	g_0^2=\frac{1-Z}{1+Z},\qquad
	Z=\frac{a_1^2-a_2^2+\pa{b_1-b_2}^2}{CD}
	\in(0,1),\qquad
	\sqrt{1-Z^2}=\frac{2a_1\pa{b_1-b_2}}{CD}
	\in(0,1).
\end{align}
We now invoke the results in \S342 of Ref.~\cite{Byrd1954} (though note that we use different conventions for the elliptic integrals).  After correcting a typo in the auxiliary formula \S361.64 (the second square root in the denominator of $f_2$ should only include $\sqrt{1+\alpha^2}$), we find that whenever
\begin{align}
	\label{eq:ConditionsForS}
	j\in\pa{0,1},\qquad
	\alpha\in\mathbb{R},\qquad
	\varphi\in\pa{-\frac{\pi}{2}+\arctan{\alpha},\frac{\pi}{2}+\arctan{\alpha}},
\end{align}
so that $\pa{1+\alpha^2}\pa{1-j+\alpha^2}>0$ is automatically satisfied, then
\begin{align}
	S_1\pa{\alpha;\varphi\big|j}&=\int_0^{F(\varphi|j)}\frac{\ed u}{1+\alpha\tn(u|j)}
	=\int_0^\varphi\frac{\ed t}{\pa{1+\alpha\tan{t}}\sqrt{1-j\sin^2{t}}}\\
	&=\frac{1}{1+\alpha^2}\br{F\pa{\varphi\big|j}+\alpha^2\Pi\pa{\left.1+\alpha^2;\varphi\right|j}-\alpha f_2},\\
	S_2\pa{\alpha;\varphi|j}&=\int_0^{F(\varphi|j)}\frac{\ed u}{\br{1+\alpha\tn(u|j)}^2}
	=\int_0^\varphi\frac{\ed t}{\pa{1+\alpha\tan{t}}^2\sqrt{1-j\sin^2{t}}}\\
	&=-\frac{1}{\pa{1+\alpha^2}\pa{1-j+\alpha^2}}\br{\pa{1-j}F\pa{\varphi\big|j}+\alpha^2E\pa{\varphi\big|j}+\frac{\alpha^2\sqrt{1-j\sin^2{\varphi}}\pa{\alpha-\tan{\varphi}}}{1+\alpha\tan{\varphi}}-\alpha^3}\nonumber\\
	&\qquad
	+\pa{\frac{1}{1+\alpha^2}+\frac{1-j}{1-j+\alpha^2}}S_1\pa{\alpha;\varphi\big|j},\\
	f_2&=\frac{p_2}{2}\log\ab{\frac{1-p_2}{1+p_2}\,\frac{1+p_2\sqrt{1-j\sin^2{\varphi}}}{1-p_2\sqrt{1-j\sin^2{\varphi}}}}
	\ge0,\qquad
	p_2=\sqrt{\frac{1+\alpha^2}{1-j+\alpha^2}}
	>0.
\end{align}

Finally, we also introduce the parameters
\begin{align}
	g_\pm=\frac{g_0x_4(r_\pm)-1}{g_0+x_4(r_\pm)}.
\end{align}
Since the conditions \eqref{eq:ConditionsForS} are satisfied by $j=k_4$, $\alpha=g_{0,\pm}$, and $\varphi=\arctan{x_4(r)}+\arctan{g_0}$, the antiderivatives
\begin{align}
	\label{eq:IrCase4}
	\mathcal{I}_0&=F^{(4)}(r),\\
	\mathcal{I}_1&=\pa{\frac{a_2}{g_0}-b_1}F^{(4)}(r)-\Pi_1^{(4)}(r),\\
	\mathcal{I}_2&=\pa{\frac{a_2}{g_0}-b_1}^2F^{(4)}(r)-2\pa{\frac{a_2}{g_0}-b_1}\Pi_1^{(4)}(r)+\Pi_2^{(4)}(r),\\
	\mathcal{I}_\pm&=\frac{g_0}{a_2\br{1-g_0x_4(r_\pm)}}\br{F^{(4)}(r)-\frac{2}{C+D}\pa{\frac{1+g_0^2}{g_0\br{g_0+x_4(r_\pm)}}}S_1\pa{g_\pm;\arctan{x_4(r)}+\arctan{g_0}\Big|k_4}},
\end{align}
are real and smooth, with\footnote{$F^{(4)}>0$ and $\Pi_2^{(4)}>0$ because $F(\varphi|j)\ge0$ and $S_2(\alpha;\varphi|j)\ge0$ whenever $\varphi\in\br{0,\frac{\pi}{2}+\arctan{\alpha}}$, $j\in\pa{0,1}$, and $\alpha>0$ (in the case of $S_2$, this is manifest from the integrand).  On the other hand, $S_1$ (and therefore $\Pi_1^{(4)}>0$) can in principle be negative as $r\to\infty$ and $x_4(r)\to\frac{\pi}{2}$.}
\begin{align}
	\label{eq:F4}
	F^{(4)}(r)&=\frac{2}{C+D}F\pa{\arctan{x_4(r)}+\arctan{g_0}\Big|k_4}
	>0,\\
	\Pi_\ell^{(4)}(r)&=\frac{2}{C+D}\br{\frac{a_2}{g_0}\pa{1+g_0^2}}^\ell S_\ell\pa{g_0;\arctan{x_4(r)}+\arctan{g_0}\Big|k_4}.
\end{align}

\subsubsection{Inversion for \texorpdfstring{$r_o(\tau)$}{radial motion}}

Before a turning point is reached, the path integral $I_r=\tau$ beginning from $r=r_s$ is still given by Eq.~\eqref{eq:tauIr}, which in this case can be inverted using the Jacobi elliptic tangent function $\tn(\varphi|k)=\sn(\varphi|k)/\cn(\varphi|k)$.  This function satisfies
\begin{align}
	\label{eq:JacobiSC}
	\tn\pa{F(\arctan{\varphi}|k)}=\varphi,
\end{align}
and is odd in its first argument, $\tn(-\varphi|k)=\tn(\varphi|k)$.  Recalling that $\mathcal{I}_r=\mathcal{I}_0=F^{(4)}(r)$,  Eqs.~\eqref{eq:tauIr} and \eqref{eq:F4} can be combined to give
\begin{align}
	\label{eq:X4}
	F\pa{\arctan{x_4(r_o)}+\arctan{g_0}\Big|k_4}=X_4(\tau),\qquad
	X_4(\tau)=\frac{C+D}{2}\pa{\nu_r\tau+\mathcal{I}_r^s}.
\end{align}
At this stage, we need to invoke the arctangent addition formula
\begin{align}
	\arctan{x}+\arctan{y}=\arctan\pa{\frac{x+y}{1-xy}}+n\pi,\qquad
	n\in\mathbb{Z},
\end{align}
(with the precise value of the integer $n$ depending on the range of $x,y\in\mathbb{R}$), to reexpress the last equation as
\begin{align}
	X_4(\tau)=F\pa{\arctan\left.\pa{\frac{g_0+x_4(r_o)}{1-g_0x_4(r_o)}}+n\pi\right|k_4}
	=F\pa{\arctan\left.\pa{\frac{g_0+x_4(r_o)}{1-g_0x_4(r_o)}}\right|k_4}+2nK(k_4),
\end{align}
where the last step follows from the quasiperiodicity property $F(\varphi+n\pi|k)=F(\varphi|k)+2nK(k)$ of the elliptic integral of the first kind [see Eqs.~\eqref{eq:QuasiPeriodicity}].  Thus, we have established that
\begin{align}
	F\pa{\arctan\left.\pa{\frac{g_0+x_4(r_o)}{1-g_0x_4(r_o)}}\right|k_4}=X_4(\tau)-2nK(k_4),
\end{align}
for some integer $n\in\mathbb{Z}$.  Applying the formula \eqref{eq:JacobiSC}, it then results that
\begin{align}
	\frac{g_0+x_4(r_o)}{1-g_0x_4(r_o)}=\tn\pa{X_4(\tau)-2nK(k_4)\Big|k_4}
	=\tn\pa{X_4(\tau)\big|k_4},
\end{align}
where in the last step, we used the periodicity property $\tn(\varphi+2K(k)|k)=\tn(\varphi|k)$.  Solving for $r_o$ using Eq.~\eqref{eq:x4}, we then find
\begin{align}
	\label{eq:Inversion4}
	r_o^{(4)}(\tau)=-a_2\br{\frac{g_0-\tn\pa{X_4(\tau)\big|k_4}}{1+g_0\tn\pa{X_4(\tau)\big|k_4}}}-b_1,
\end{align}
where we write $r_o^{(4)}$ to emphasize that this formula for $r_o(\tau)$ was derived in case (4), even though it will extend to the other cases with little modification.  This trajectory never encounters a turning point outside the horizon, and hence Eq.~\eqref{eq:Inversion4} is the unique solution for $r_o(\tau)$ with initial conditions $r_o(0)=r_s$ and $\sign\br{r_o'(0)}=\nu_r$, which is manifestly real in case (4).

\subsubsection{Path integrals as a function of Mino time}

We have $I_r=I_0=\tau$ by definition, and the other path integrals may be expressed in terms of Mino time $\tau$ using the same method as usual.  Since there are no turning points, we once again have Eq.~\eqref{eq:DefiniteI}, where the antiderivatives $\mathcal{I}_i$ depend on $r$ primarily through the combination $\arctan{x_4(r)}+\arctan{g_0}$.  We now invoke the inversion formula
\begin{align}
	\am(F(\varphi|k)|k)=\varphi.
\end{align}
Applying it to Eq.~\eqref{eq:X4} extends $\arctan{x_4(r)}+\arctan{g_0}$ to the (monotonically increasing in $\tau$) amplitude
\begin{align}
	\label{eq:x4Amplitude}
	\arctan{x_4(r)}+\arctan{g_0}=\am\pa{X_4(\tau)\big|k},
\end{align}
where $X_4(\tau)$ is as defined in Eq.~\eqref{eq:X4}.  Plugging the extension \eqref{eq:x4Amplitude} into Eq.~\eqref{eq:DefiniteI} as needed, we then find
\begin{align}
	I_1&=\pa{\frac{a_2}{g_0}-b_1}\tau-\Pi_{1,\tau}^{(4)},\\
	I_2&=\pa{\frac{a_2}{g_0}-b_1}^2\tau-2\pa{\frac{a_2}{g_0}-b_1}\Pi_{1,\tau}^{(4)}+\Pi_{2,\tau}^{(4)},\\
	I_\pm&=\frac{g_0}{a_2\br{1-g_0x_4(r_\pm)}}\pa{\tau-\Pi_{\pm,\tau}^{(4)}},
\end{align}
with
\begin{align}
	\Pi_{\ell,\tau}^{(4)}&=\frac{2\nu_r}{C+D}\br{\frac{a_2}{g_0}\pa{1+g_0^2}}^\ell\br{S_\ell\pa{g_0;\am\pa{X_4(\tau)\big|k_4}\Big|k_4}-S_\ell\pa{g_0;\arctan{x_4(r_s)}+\arctan{g_0}\Big|k_4}},\\
	\Pi_{\pm,\tau}^{(4)}&=\frac{2\nu_r}{C+D}\pa{\frac{1+g_0^2}{g_0\br{g_0+x_4(r_\pm)}}}\br{S_1\pa{g_\pm;\am\pa{X_4(\tau)\big|k_4}\Big|k_4}-S_1\pa{g_\pm;\arctan{x_4(r_s)}+\arctan{g_0}\Big|k_4}}.
\end{align}
These functions of Mino time vanish at $\tau=0$ by construction.  Since a turning point never occurs, these formulas give the full path integrals \eqref{eq:EllipticIntegralsBasis}.

\subsection{Unified inversion formula}

Although the inversion formulas $r_o^{(i)}(\tau)$ derived in the four cases (1)-(4) appear superficially different, their expressions \eqref{eq:Inversion1}, \eqref{eq:Inversion2}, \eqref{eq:Inversion3}, and \eqref{eq:Inversion4} are in fact mathematically equivalent after certain sign flips in $\nu_r$.  Thus, either one may be used across all four cases (with appropriate sign flips), even though it is only manifestly real in its own domain of definition.

To prove this, it suffices to show that for each $i\in\cu{1,2,3,4}$, the function $r_o(\tau)=r_o^{(i)}(\tau)$ always satisfies the squared differential equation \eqref{eq:RadialODE} with the correct initial position.  One can then check the initial sign of the derivative and send $\nu_r\to-\nu_r$ if necessary.  For instance, for the inversion formula \eqref{eq:Inversion2} derived in case (2), one has
\begin{align}
	\label{eq:RadialDerivative}
	\frac{dr_o^{(2)}}{d\tau}&=\frac{r_{31}r_{41}r_{43}\sqrt{r_{31}r_{42}}\sn\pa{X_2(\tau)\big|k}\cn\pa{X_2(\tau)\big|k}\dn\pa{X_2(\tau)\big|k}}{\br{r_{31}-r_{41}\sn\pa{X_2(\tau)\big|k}}^2},\\
	\mathcal{R}\pa{r_o^{(2)}(\tau)}&=\frac{r_{31}^3r_{41}^2r_{43}^2r_{42}\sn^2\pa{X_2(\tau)\big|k}\br{1-\sn^2\pa{X_2(\tau)\big|k}}\br{1-k\sn^2\pa{X_2(\tau)\big|k}}}{\br{r_{31}-r_{41}\sn\pa{X_2(\tau)\big|k}}^4}.
\end{align}
These expressions manifestly satisfy Eq.~\eqref{eq:RadialODE}, in light of the elliptic identities \eqref{eq:JacobiSNCN} and \eqref{eq:JacobiSNDN}.  Hence, $r_o^{(2)}(\tau)$ remains valid in all cases, up to a possible sign flip in $\nu_r$.\footnote{An almost identical derivation goes through for $r_o^{(1)}(\tau)$ and $r_o^{(3)}(\tau)$, so we omit it here.  We did not analytically complete the proof for $r_o^{(4)}(\tau)$, which seems to require complicated elliptic identities, but we have numerically checked that it also remains valid in all cases.  On the other hand, it is easy to explicitly check that $r_o^{(i)}(0)=r_s$ always holds for all $i\in\cu{1,2,3,4}$.}  We can fix the sign ambiguity in $\nu_r$ by careful examination of Eq.~\eqref{eq:RadialDerivative}.  By construction, the initial sign of the radial momentum in case (2) is $\sign\pa{p_s^r}=\nu_r$.  An explicit but tedious computation reveals this to still hold in cases (3) and (4); on the other hand, a sign flip is required for case (1), in which $\sign\pa{p_s^r}=-\nu_r$.  Thus a unified inversion formula holding in all cases is
\begin{align}
	r_o^{(2)}(\tau)=\frac{r_4r_{31}-r_3r_{41}\sn^2\pa{X_2(\tau)\big|k}}{r_{31}-r_{41}\sn^2\pa{X_2(\tau)\big|k}},\qquad
	X_2(\tau)=\frac{\sqrt{r_{31}r_{42}}}{2}\pa{\tau+\alpha\nu_r\mathcal{I}_r^s}.
\end{align}
where $\alpha=-1$ in case (1) and $\alpha=+1$ in all other cases (2)-(4).  Here, $\mathcal{I}_r^s$ refers to the case (2) antiderivative \eqref{eq:IrCase2}.

%

\bibliography{KerrRing}
\bibliographystyle{utphys}

\end{document}